\documentclass[useAMS,usenatbib]{mn2e}
 \usepackage{lscape}
\usepackage{natbib}
\usepackage{epsfig}
\usepackage{rotating}
\usepackage{amsmath}
\usepackage{tabu}
\usepackage{adjustbox}
\usepackage{graphicx}
\usepackage{float}
\usepackage{placeins}
\usepackage{footnote}
\usepackage{tablefootnote}

\def\gtrsim{\,\lower2truept\hbox{${>\atop\hbox{\raise4truept\hbox{$\sim$}}}$}\,}
\title[ALMA Band 3 polarimetric follow-up of a complete sample]{ALMA Band 3 polarimetric follow-up of a complete sample of faint PACO sources}
\author[V. Galluzzi et al.]{\parbox[t]{\textwidth}{
V.~Galluzzi$^{1,2}$\thanks{E-mail: vincenzo.galluzzi@inaf.it (VG)}, G.~Puglisi$^{3,4}$, S.~Burkutean$^{2}$, E.~Liuzzo$^{2}$, M.~Bonato$^{2}$, M.~Massardi$^{2}$\thanks{E-mail: massardi@ira.inaf.it (MM)}, R.~Paladino$^{2}$, L.~Gregorini$^{2}$,  R.~Ricci$^{2}$, T.~Trombetti$^{2,5}$, L.~Toffolatti$^{6,7}$, C.~Burigana$^{2,8,9}$,
A.~Bonaldi$^{10}, $ L.~Bonavera$^{6}$, V.~Casasola$^{2,14}$, G.~De Zotti$^{11}$, R.~D.~Ekers$^{12,13}$, S.~di Serego Alighieri$^{14}$, M.~L\'opez-Caniego$^{15}$ and M.~Tucci$^{16}$}
\vspace*{8pt}\\
$^{1}$INAF, Osservatorio Astronomico di Trieste, via Gian Battista Tiepolo 11, I-34143 Trieste, Italy\\
$^{2}$INAF, Istituto di Radioastronomia, via Piero Gobetti 101, I-40129 Bologna, Italy\\
$^{3}$SISSA, via Bonomea 265, I-34136 Trieste, Italy\\
$^{4}$INFN-Sezione di Trieste, via Valerio 2, I-34127 Trieste, Italy\\
$^{5}$INFN-Sezione di Ferrara, via Giuseppe Saragat 1, I-44122, Ferrara, Italy\\
$^{6}$Departamento de F\'isica Universidad de Oviedo, C. Federico Garc\'ia Lorca 18, E-33007 Oviedo, Spain\\
$^{7}$INAF-OAS Bologna, via Piero Gobetti 93/2, I-40129 Bologna, Italy\\
$^{8}$Dipartimento di Fisica e Scienze della Terra, Universit\`a degli Studi di Ferrara, via Giuseppe Saragat 1, I-44100 Ferrara, Italy\\
$^{9}$INFN-Sezione di Bologna, via Irnerio 46, I-40126 Bologna, Italy\\ 
$^{10}$SKA Organization, Jodrell Bank, Lower Whitington, Macclesfield, SK11 9DL, UK\\
$^{11}$INAF, Osservatorio Astronomico di Padova, Vicolo dell'Osservatorio 5, I-35122 Padova, Italy\\
$^{12}$CSIRO Astronomy and Space Science, PO Box 76, Epping, NSW 1710, Australia\\
$^{13}$International Centre for Radio Astronomy Research, Curtin University, Bentley, WA 6102, Australia\\\
$^{14}$INAF - Osservatorio Astrofisico di Arcetri, Largo Enrico Fermi 5, I-50125 Firenze, Italy\\
$^{15}$European Space Agency, ESAC, Camino bajo del Castillo, s/n, Urbanizaci\'{o}n Villafranca del Castillo,\\
\;\;\,Villanueva de la Ca\~{n}ada, , E-28692 Madrid, Spain\\
$^{16}$D\'epartement de Physique Th\'eorique and Center for Astroparticle Physics (CAP), University of Geneva,\\\;\;\, 24 quai Ernest Ansermet, CH-1211 Geneva, Switzerland}
\begin{document}

\date{}

%\pagerange{\pageref{firstpage}--\pageref{lastpage}}
\pubyear{2010}

\maketitle

\label{firstpage}

% Abstract of the paper
\begin{abstract}
We present Atacama Large Millimeter/submillimiter Array (ALMA) high sensitivity ($\sigma_P \simeq 0.4\,$mJy) polarimetric
observations at $97.5\,$GHz (Band 3) of a complete sample of $32$
extragalactic radio sources drawn from the faint \textit{Planck}-ATCA Co-eval Observations (PACO) sample ($b<-75^\circ$,
compact sources brighter than $200\,$mJy at $20\,$GHz). We achieved a
detection rate of $~97\%$ at $3\,\sigma$ (only $1$ non-detection). We complement
these observations with new Australia Telescope Compact Array (ATCA) data between $2.1$ and $35\,$GHz obtained
within a few months and with data published in earlier papers from our
collaboration. Adding the co-eval GaLactic and Extragalactic All-sky Murchison widefield array (GLEAM) survey detections between $70\,$ and
$230\,$MHz for our sources, we present spectra over more than $3$ decades in
frequency in total intensity and over about $1.7$ decades in polarization. The spectra of our sources are smooth over the whole frequency range, with no
sign of dust emission from the host galaxy at mm wavelengths nor of a sharp
high frequency decline due, for example, to electron ageing. We do however find
indications of multiple emitting components and present a classification
based on the number of detected components. We analyze the polarization
fraction behaviour and distributions up to $97\,$GHz for different source
classes. Source counts in polarization are presented at $95\,$GHz.
%Their implications for CMB experiments are addressed in the companion paper by Puglisi et al. %We
%also discuss a few relevant cases (AT20GJ040848-750720 and PKS0521-365, i.e.
%the leakage calibrator) resolved in our ALMA dataset.
\end{abstract}

% Select between one and six entries from the list of approved keywords.
% Don't make up new ones.
\begin{keywords}
galaxies: active -- radio continuum: galaxies -- galaxies: statistics.
\end{keywords}
\vspace*{2.0cm}
%%%%%%%%%%%%%%%%% BODY OF PAPER %%%%%%%%%%%%%%%%%%

\section{Introduction}

The most commonly used model for the spectral energy distribution (SED) of
blazars, i.e. compact, radio loud active galactic nuclei (AGNs), is a leptonic, one-zone
model, where the emission originates in a single component
\citep[][]{Bottcher2012}. The SEDs typically consist of two broad-band bumps:
the one at lower frequencies is attributed to synchrotron radiation while the
second, peaking at $\gamma$-ray energies, is attributed to inverse Compton.

The one-zone model is generally found to provide an adequate approximation
primarily because of the limited observational characterization of the
synchrotron SED, with fragmentary data over a limited frequency range. However,
the synchrotron emission is originated by relativistic jets, and Very Long Baseline Interferometry (VLBI) images
show multiple knots often called ``components'' of the jet. The standard model
interprets the knots as due to shocks that enhance the local synchrotron emission.

The spectrum is explained as the result of the superposition of different
synchrotron self-absorbed components in a conical geometry
\citep[][]{Marscher1996}. The synchrotron self-absorption optical depth scales
as $\tau_{\rm sync}\propto B_\bot^{(p+2)/2}\nu^{-(p+4)/2}$ where $B_\bot$ is
the magnetic field component perpendicular to the electron velocity and $p$ is
spectral index of the energy distribution of relativistic electrons (typically,
$p\simeq 2.5$). Thus, $\tau_{\rm sync}$ increases along the jet towards the
nucleus as the magnetic field intensity and its ordering increases. At the same time, it is
strongly frequency dependent: as the observing frequency increases, the
emission becomes detectable at progressively smaller distances from the central
engine.

Thus, the millimeter-wave emission provides information on the innermost regions
of the jets, close to the active nucleus, where it is optically thin, while the
emission at longer wavelengths is affected by self-absorption \citep{Jorstad2007,Agudo2014}. Interestingly, \citet{LeonTavares2011}, \citet{Jorstad2013} and
\citet{Ramakrishnan2016} found a significant correlation between simultaneous
$\gamma$-ray fluxes and millimeter-wave flux  densities of
flat spectrum radio quasars (FSRQs), especially with high optical
polarization. The strongest $\gamma$-ray flares were found to occur during the
rising/peaking stages of millimeter flares. This suggests that the $\gamma$-ray
flares originate in the millimeter-wave emitting regions of these sources.

Polarization carries information on the magnetic field configuration (geometry
and degree of order). In the shocked regions in the jet,
the magnetic field is compressed. The compression makes it effectively more
ordered, increasing the polarization degree \citep[see, e.g., ][]{Hughes1989}.
Multifrequency polarimetry is therefore a key indicator of the physical
conditions in a jet.

%{\bf The ``shock-in-jet'' model assumes that a shock wave in propagating through a conical jet accelerates particles at the shock front up to relativistic speeds. Then, particles lose their energy through different stages, namely Compton, synchrotron and abiabatic losses. Flaring activitiy emerging from observed light curves at different frequencies and quantified in terms of amplitudes, time scales and cross-band delays has been extensively explained in this scenario \citep[e.g.,][]{Fromm2015,Fuhrmann2016}.}
The origin of the observed strong variability in the synchrotron emission of blazars is still debated. The ``shock-in-jet'' model \citep[][]{MarscherGear1985,Hughes1985} was shown to provide a promising framework to account for the frequency dependencies of variability amplitudes and time scales \citep[e.g.,][]{Fromm2015,Fuhrmann2016}. According to this model, a shock wave propagates through a conical jet. Knots are interpreted as the bright downstream regions of such flow structure. Particles are acceleratee to relativistic energies at the shock front and then loose energy via Compton scattering, synchrotron emission and adiabatic losses. The variation of the Doppler factor along the jet has a major role in determining the frequency dependence of the variability parameters.

Polarization variability provides particularly useful clues to modelling \citep{Hughes1989}. While in the quiescent phase blazars are polarized at a few percent level, individual knots can be highly polarized. This implies that the overall magnetic field is highly turbulent, the field in regions responsible for the outbursts is much more ordered. Hence the frequency dependence of the polarized emission is a powerful tool to identify emission regions that would be otherwise unresolved.
 However, evidences of multi-component contributions to the
synchrotron SED are still limited, although the situation has been improving in
recent years \citep[e.g.,][]{PlanckCollaborationXV2011, Cutini2014}.

Our group has been carrying out a long-term programme of multi-frequency
observations with the Australia Telescope Compact Array (ATCA) of the \textit{Planck}--ATCA Co-eval Observations (PACO) sample. The PACO project \citep[][and
 references therein]{Massardi2016} observed $482$ Australia Telescope $20\,$GHz survey (AT20G) extragalactic sources (at Galactic latitude $|b|> 5^\circ$ and outside a $5^\circ$ radius circle around the Large Magellanic Cloud). Of these, $344$ objects constitute three partially overlapping sub-samples, selected for different purposes: the ``faint sample'' comprises $159$ sources with $S_{20\,{\rm GHz}}>200\,$mJy with $3\,{\rm h}<{\rm RA}<9\,{\rm h}$ and $\delta < -30^\circ$, and allowed us to characterize radio source spectra below the sensitivity of the \textit{Planck} satellite over an area near to the southern ecliptic pole (where \textit{Planck} sensitivity was maximal); the ``bright sample'', namely the $189$ sources with $S_{20\,{\rm GHz}}> 500\,$mJy and $\delta < -30^\circ$, and the ``spectrally-selected'' one, i.e. the $69$ sources  with $S_{20\,{\rm GHz}}>200\,$mJy (over the whole southern sky) classified as inverted- or upturning-spectrum by \citet{Massardi2011b}.

\citet{Galluzzi2018} have presented high
sensitivity polarimetric observations in seven bands, from $2.1$ to $38\,$GHz, of $104$ compact extragalactic radio sources drawn from the faint PACO sub-sample, i.e. brighter
than 200\,mJy at 20\,GHz. Combining these results with the GaLactic and
Extra-galactic All-sky Murchison widefield array (GLEAM) survey data at $20$ frequencies
between $72$ and $231\,$MHz \citep{HurleyWalker2017}, it was found that about $90\%$ of
their sources showed clear indications of at least two emission components. The
broad frequency coverage and the polarimetry proved to be essential to reach
this conclusion: total intensity data from $5.5$ to $38\,$GHz could be interpreted
in terms of a single emission component \citep{Galluzzi2017}.

In this paper, we extend the frequency coverage in total and polarized intensity
of a complete sub-sample of 32 sources, drawn from the \citet{Galluzzi2018}
sample, by means of high sensitivity observations  with the Atacama Large
Millimeter/submillimiter Array (ALMA) at $97.5\,$GHz (Band 3).

Apart from providing information on the physics of inner regions of
relativistic jets, mm-wave polarimetric observations have two other important
astrophysical applications.

Radio sources are the dominant contaminants of Cosmic Microwave Background
(CMB) maps on small scales down to mm wavelengths. An accurate characterization
of their polarization properties is especially crucial for attempts to measure
the primordial $B$-mode polarization down to values of the tensor to scalar
ratios $r\sim ~0.001$. The accurate simulations by \citet{Remazeilles2018} have
shown that, at these values of $r$,  unresolved polarized point sources can be
the dominant foreground contaminant over a broad range of angular scales
(multipoles $\ell \gtrsim 50$). These results have been confirmed by \citet{Puglisi2018} who exploited the state-of-the-art data sets of polarized point sources over the $1.4-217\,$GHz frequency range  (including the distribution of polarization fractions presented in this paper), in order to forecast extragalactic radio sources contamination of the CMB B-mode angular power spectrum, with reference to some existing or planned ground-based or space-borne CMB facilities (e.g. QUIJOTE\footnote{Q-U-I JOint TEnerife.}, LiteBIRD\footnote{Lite satellite for the studies of B-mode polarization and Inflation from cosmic Background Radiation Detection.} and CORE\footnote{Cosmic ORigin Explorer.}): since the other important point source population in the frequency range of CMB experiments, dusty galaxies, is believed to be very weakly polarized, radio sources are expected to dominate
small-scale polarization fluctuations up to $\simeq 150\,$GHz.

%both in total intensity and in polarization and affect its angular power
%spectrum up to $\sim 30\,$arcmin. Due to the lack of observations (especially
%n polarization) at high frequency ($> 20\,$GHz), it is unsafe ({\it cf.}
%Huffenberger et al. 2015) to infer properties of radio sources at frequencies
%suitable for cosmological studies (between $70$ and $160\,$GHz) by
%extrapolating from low frequency observations. Furthermore, extragalactic radio
%sources still constitute the best calibrators both of the polarized intensity
%and of the polarization angle for CMB experiments. The systematic errors due to
%inaccuracies in the calibration of the polarization angle are becoming the
%limiting factor for CMB polarization experiments (e.g. Kaufman et al. 2016).

A polarimetric AGN catalogue at millimeter wavelengths is also necessary for the calibration of CMB maps.
Furthermore, theoretical studies have examined the possible existence of terms in the Lagrangian density that can violate the Einstein Equivalence Principle (EEP), the Lorentz invariance or the CPT invariance \citep[e.g. see][]{Ni2010}. These terms would produce a rotation of the polarization angle along the propagation of the electromagnetic wave: this is the so-called Cosmic Polarization Rotation (CPR). The best upper limits on the CPR from CMB experiments and observations of astrophysical objects in optical or radio bands are around $1^\circ$ \citep{diSeregoAlighieri2015} and are limited by the calibration accuracy of the zero point polarization angle. Finding bright point-like objects with at least a few percent polarization fraction at high frequencies (where Faraday rotation is typically negligible) and with stable polarimetric properties (in particular, a constant polarization angle) at least on a few years timescale may help in constraining this effect on a sub-degree scale. With our multi-frequency and multi-epoch ATCA observations corroborated by the present ALMA follow-up at $97.5\,$GHz, we identify some potential candidates for CPR studies calibration.

The paper is organised as follows. In Section\,\ref{sec:obs}  we  present the
observational campaigns. In Section\,\ref{sec:datareduc} we briefly describe
the data reduction and flux density extraction. In Section\,\ref{sec:dataanaly}
we discuss the data analysis, the spectral behaviour and the polarimetric
properties of sources.
%In Section\,\ref{sec:variability} we update the variability study presented in \citet{Galluzzi2017}.
In Section\,\ref{sec:soucou100ghz} we present the source counts in polarized flux
density at $95\,$GHz obtained by convolving the total intensity differential
source counts with the observed polarization fraction distribution. In
Section\,\ref{sec:pecobj} we discuss peculiar objects, such as the Fanaroff-Riley Class II (FR-II)
object AT20G0408-750528 and the blazar PKS0521-365 (our leakage calibrator),
for which a more extended multi-frequency and multi-epoch investigation at VLBI
resolution will be presented in a forthcoming paper (Liuzzo et al., in
preparation). We also identify some potential CPR calibrators. Finally, in Section\,\ref{sec:discusseconcl} we draw our
conclusions.

\section{ALMA Observations}
\label{sec:obs}

%-----------------------------Figure Start---------------------------
%\begin{landscape}
\begin{figure}
	%\vspace{-2.5cm}
	\centering
	\includegraphics[scale=0.33,trim={0.45cm 0.4cm 0 0},clip]{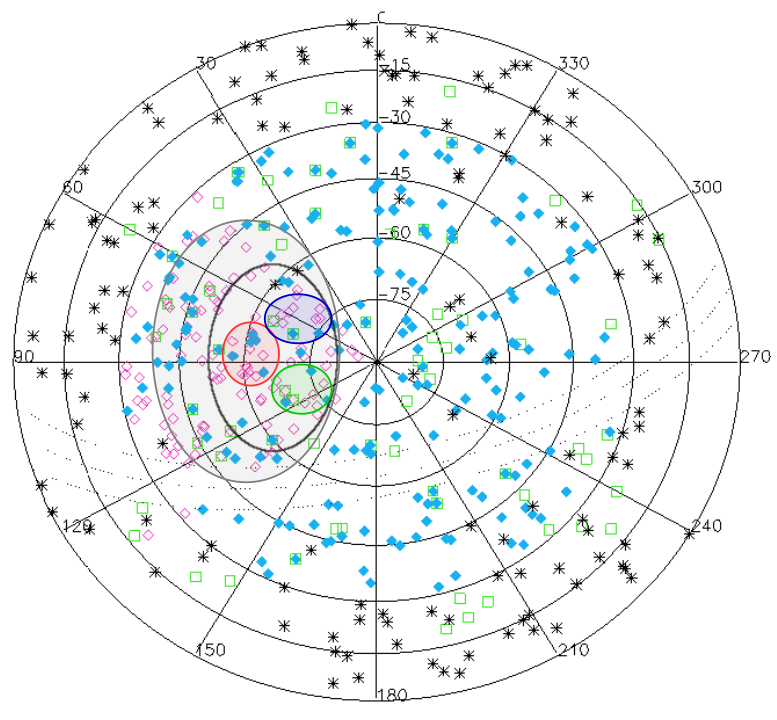}
	%\vspace{-0.85cm}
	\caption{Polar equal-area projection map showing the positions of PACO sources (RA and Dec are in degrees). The faint PACO sources are shown as open pink diamonds, the bright PACO ones as filled blue diamonds. Green squares are for the spectrally-selected PACO sample and black asterisks are for blazars and Australia Telescope (AT) calibrators, respectively. The dotted lines indicate the Galactic plane and mark the area with Galactic latitude $|b| < 5^\circ$ \citep[\textit{cf.}][]{Massardi2016}. The black and the grey ellipses enclose the samples investigated by \citet{Galluzzi2017} and \citet{Galluzzi2018}, respectively. The smaller red, green and blue ellipses encircle the three Science Goals of ALMA observations.}
	\label{fig:PolObsMap}
\end{figure}
%\end{landscape}
%-----------------------------Figure End------------------------------

The observations  were carried out with ALMA (Cycle 3, Project ID: 2015.1.01522.S,
PI: Galluzzi) on 24th August, 22th and 27th September $2016$, in four
$2\,$GHz-wide spectral bands centered at $90.5\,$, $92.5$, $102.5$ and
$104.5\,$GHz, respectively, using $39$ antennas in a compact configuration (baseline range $118 - 1318\,$m, corresponding to resolutions of $4.8-0.3\,$arcsec at 97.5 GHz).

We observed a complete sample of $32$ objects drawn from the faint PACO sample
($S_{20\,GHz}> 200\,$mJy) in three circular regions at $b<-75^\circ$ (each with $\sim 10^\circ$
diameter) that, altogether, contain $\simeq 60\%$ of the 53 sources observed by
\citet{Galluzzi2017}. The three regions were selected in order to optimize the
use of ALMA time, maximizing the sample size with the smallest possible number
of Science Goals (SGs; see Fig.~\ref{fig:PolObsMap}), where the term ``Science Goal'' indicates a small group of sources which share the same spectral and sensitivity requirements, and the same calibration.
%In fact, each ALMA SG in Cycle 3 should include only objects within $10^\circ$ from each other, so that they could share the same calibration.
The latter requires at least $3\,$hr of observations for each polarimetric SG, in which observations of the target are interleaved with those of the polarization calibrator, to achieve the adeguate parallactic angle coverage for the computation of polarimetric ``leakage'' (D-terms).

%We requested a sensitivity of $\sim 30\,\mu$Jy. This sensitivity level was
%estimated extrapolating to 100\,GHz the spectral fits for each source obtained
%by \citet{Galluzzi2017} on the basis of their previous ATCA total intensity
%observations in the range 5.5--38\,GHz, and conservatively assuming a
%polarization fraction of 1\%. In some cases we reached a sensitivity of
%$20\,\mu$Jy, since objects are observed three times (once per execution block),
%resulting also in a better coverage of the $uv$ plane, thanks to the $39$ ALMA
%$12\,$m antennas available in Cycle 3 and to the $8\,$GHz continuum bandwidth
%for each polarization.
% A first execution was not enough to reach the requested sensitivity for our $3$ SGs so that each scheduling block was executed twice reaching sensitivities of the order of ... and a better uv coverage, thanks to the $39$ ALMA antennas available in Cycle 3 and to the 8GHz continuum bandwidth for each polarization.

This allowed us to get in linear polarization a $3\,\sigma$ detection rate of $97\%$ (only one
non-detection) and a $5\,\sigma$ detection rate of $\simeq 94\%$ (only two
non-detections). The median significance of detections is $\simeq 10\,\sigma$.

The sources were unresolved by ATCA at all frequencies (up to $38\,$GHz). Our
ALMA observations achieved a resolution of $\simeq 0.2\,$arcsec, a factor
$\simeq 25$ higher than that of ATCA observations at $38\,$GHz. The possibility
that some sources might be resolved by ALMA was considered in our flux density
estimation approach and in some of the analyses described in the following
sections.

\section{Data reduction}
\label{sec:datareduc} % used for referring to this section from elsewhere
ALMA data were calibrated by using the Common Astronomical Software Applications (CASA) version 4.7.0, following the current standard calibration scheme reported in \citet{Nagai2016} and the CASA Guide\footnote{https://casaguides.nrao.edu/index.php/Main\_Page}. %However,
%an expert PI might decide to recalibrate data, e.g. adopting a self calibration
%approach to improve the signal-to-noise (S/N) ratio, hence reducing the rms in
%the calibrated image.

In Table~\ref{tab:ALMAObsCal} we report the list of the calibrators visited
during the ALMA observations. The ALMA data reduction consists of two steps: the
first one corrects only the parallel hands products, i.e. XX and YY and the
second one (needed in case of polarimetry) addresses the cross products XY and
YX, and the refinement of XX and YY gains.
%An intermediate step, which prepares the data coming from the first step for the second one, is performed by the task CONCAT, which first orders the visibilities of the same object in different scans and then concatenates them.

Since almost all the objects are point-like, $I$, $Q$ and $U$ Stokes flux densities
are extracted from the corresponding maps (obtained with a natural weighting) by modelling the emission with a 2D
Gaussian (whose widths are of the order of the FWHM of the synthesized beam, i.e. $0.3\,$arcsec) and deriving the integrated flux densities. Whenever the fit fails
because the source is too faint to be detected in Stokes $Q$ and $U$, we
consider the central peak in the image. The flux density extraction for resolved
objects is addressed in Section\,\ref{sec:pecobj}.

In the Tab.~\ref{tab:ALMAObsSum} we report details about the array configuration, the minimum and maximum angular scales, the time on source and the sensitivity achieved for each SG.

\begin{table}
\caption{List of the calibrators visited during our ALMA observations. The second column reports the Science Goal (SG) observed in a given epoch.}
\label{tab:ALMAObsCal}
\begin{adjustbox}{max width=\columnwidth}
\begin{tabular}{lccccc}
\hline
Epoch & SG & Bandpass & Flux & Phase & Leakage\\
24/08/16 & 1 & J0635-7516 & J0519-4546 & J0715-6829 & J0538-4405\\
22/09/16 & 3 & J0635-7516 & J0519-4546 & J0440-6952 & J0522-3627\\
27/09/16 & 2 & J0635-7516 & J0519-4546 & J0715-6829 & J0538-4405\\
\hline
\end{tabular}
\end{adjustbox}
\end{table}

\begin{table}
\caption{\small ALMA array configuration, minimum and maximum angular scales, time on source and sensitivity for each SG of our observations. }
\begin{tabular}{cccccc}
\hline
SG & Array & min.-max.& time on& sens.\\
& conf. & scale ('') & source (min) & ($\mu$Jy)\\ 
\hline
1 & C40-6 & $0.4-4.8$ & 5.04 & 40\\
3 & C40-6 & $0.2-4.8$ & 11.69 & 20\\
2 & C40-6 & $0.2-4.8$ & 11.69 & 20\\
\hline
\end{tabular}
\label{tab:ALMAObsSum}
\end{table}

%Our sources are known to exhibit linear polarization \citep[up to $\sim 10\%$;][]{Massardi2008, Massardi2013}, defined by the $Q$ and $U$ Stokes parameters. Observations of the circular polarization of extragalactic radio sources have demonstrated that it is very low, generally below $0.1-0.2\%$, at least one order of magnitude lower than the linear polarization \citep{Rayner2000}, with a quite steep behaviour with the frequency $\alpha ~0.5-1.0$. Hence, the rms $\sigma_V$ of the retrieved $V$ Stokes parameter is frequently used as a noise estimator.
During ALMA Cycle 3 Stokes V (circular polarization) was still under commissioning. Stokes V images obtained were not reliable, hence, differently from our previous works \citep{Galluzzi2017,
Galluzzi2018}, we cannot use the first-order debiasing technique. However, our
experience with high sensitivity ($0.6\,$mJy) ATCA data has shown that the
debiasing term $\sigma_V$ lowers the estimated value of the polarized flux
density by $\sim 0.01\%$, well within our assumed calibration error ($\sim
10\%$). Hence, the linearly polarized emission, $P$, can be safely estimated from the
Stokes parameters $Q$ and $U$ only:
%
%Our high sensitivity measurements have allowed us to check the soundness of this procedure. We achieved a $5\,\sigma$ detection of circular polarization, $V$, in $\sim 38\%$ of the dataset, i.e. $\sim 89\%$ of the objects are detected at least at one frequency. For only $\sim 15\%$ of detections, the circular to linear polarization ratio is $\ge 20\%$; the mean circular polarization is substantially smaller than our calibration error of the polarized flux density, which is $\simeq 10\%$ \citep{Galluzzi2017}. We demand the discussion about the circular polarization to the sub-sect.~\ref{subsec:CirPol}.
%
\begin{equation}
P=\sqrt{Q^2+U^2}\,.
\label{equ:PolFluDen}
\end{equation}
%
%The $\sigma_V^2$ term removes the noise bias on $P$ \citep[e.g.,][]{WardleKronberg1974}.\footnote{The error associated to the bias correction is negligible and will be ignored in the following.} We find that ignoring the $V$ contribution in eq.~(\ref{equ:PolFluDen}) results in a mean relative (sistematic) error of 0.01\%, which is, however, three order of magnitudes lower than the the assumed calibration error in polarization.
%
The polarization angle $\phi$ and the polarization fraction $\Pi$ (usually in
terms of a percentage) write:
\begin{eqnarray}
\phi&=&\frac{1}{2}\arctan{\left(\frac{U}{Q}\right)},\\
\Pi&=&100\cdot P/I.
\end{eqnarray}
The errors in total intensity, linear polarization flux density and position
angle were computed as in \citet{Galluzzi2017}, i.e. adopting calibration
errors added in quadrature to the statistical ones. The CASA Guide recommends to use a $10\%$ of the measured flux density for Stokes' $I$, $Q$ and $U$ and an additional $2^{\circ}$ for the instrumental error on the polarization angle. Indeed we assumed a lower error for $I$ (i.e. $7\%$) because the primary calibrator, namely the core of Pictor A (AT20GJ051949-454643), is found to be stable within $\sim 2\%$ both at $91.5$ and $103.5\,$GHz during the one month period before and after our observations.
%Flux densities are estimated via the IMFIT task in the image plane: since our objects are not resolved according to the ALMA synthesized beam, we use the peak (that might be positive or negative, in case of Q and U) of the images in all the Stokes. Just in case the object is resolved, we estimate the total flux density integrating onto the emission region, as in case of AT20GJ0408-7507. For this source we also have a resolved structure in Q and U. These, differently from I are intrinsic vector quantities, hence the eventual integration should to have into account also the orientation of each pixel. However, we simply have three compact regions whose flux densities could be easily recovered from the peak of image for each emitting region.

All the flux densities (total intensity and polarization), the polarization
angle and the polarization fractions are reported in Table~\ref{tab:catalogue}.
%***MM fino a qui.

\begin{table*}
\caption{\small ALMA Band 3 (central frequency: $97.5\,$GHz) observations performed at the end of August and at the end of September 2016. The table below reports: the sequential number, the AT20G name, the GLEAM counterpart, RA (in hours) and Dec (in degrees), an extension flag (``.'' for ``point-like'', ``e'' for ``extended in this observation'' and ``pe'' for ``probably extended''), the flux density in total intensity (Stokes' I), the linearly polarized flux density (P), the linear polarization fraction ($\Pi$), the polarization angle ($\phi$) and associated errors for all these quantities (``$<$'' markes a $3\,\sigma$ upper limit, while ``-'' stands for not available data). Note that we do not catalogue polarization values for the object AT20GJ040848-750720, since it is well resolved in three linearly polarized components ({\it cf.} Fig.~\ref{fig:AT20GJ040848-750720_composed}). A machine-readable version of this catalogue is available as online supplementary material.}
{\scriptsize
\begin{tabular}{lccccccccccccc}
\hline
No & (AT20G) name & (GLEAM) name & RA ($h$) & Dec ($^\circ$) & Flag & I (mJy) & err$_{\rm I}$ & $P$ (mJy) & err$_P$ & $\Pi$ ($\%$) & err$_\Pi$ & $\phi$ ($^\circ$)& err$_\phi$\\
\hline
  1 &  J032404-732047 & J032400-732039  &  3.4011192 & -73.3463898 & . & 63.42 & 4.44 & 1.90 & 0.19 & 3.00 & 0.37 & - & - \\
  2 &  J033243-724904 & J033242-724906  &  3.5453087 & -72.8180313 & . & 75.48 & 5.29 & 1.29 & 0.13 & 1.71 & 0.21 & 3.6 & 2.2\\
  3 &  J034028-670316 & J034028-670315  &  3.6744947 & -67.0546722 & . & 145.54 & 10.19 & 0.66 & 0.08 & 0.45 & 0.06 & -38.0 & 3.1\\
  4 &  J035547-664533 & J035548-664532  &  3.9299614 & -66.7593613 & . & 241.01 & 16.88 & 1.93 & 0.17 & 0.80 & 0.09 & -57.1 & 2.6\\
  5 &  J040820-654508 & J040820-654458  &  4.1390414 & -65.7522812 & pe & 24.28 & 1.70 & 2.11 & 0.16 & 8.69 & 0.90 & -63.2 & 2.8\\
  6 &  J040848-750720 & J040848-750716  &  4.1468747 & -75.1222534 & e & 106.55 & 7.46 & - & - & - & - & - & -\\
  7 &  J042506-664650 & J042507-664656  &  4.4185832 & -66.7805786 & . & 33.24 & 2.33 & 0.83 & 0.07 & 2.51 & 0.28 & -67.0 & 3.2\\
  8 &  J044047-695217 & - &   4.6799779 & -69.8715286 & . & 307.31 & 21.52 & 11.12 & 1.11 & 3.62 & 0.44 & 92.5 & 2.0\\
  9 &  J050644-610941 & J050643-610941 &  5.1122279 & -61.1614990 & pe & 370.48 & 25.95 & 5.38 & 0.53 & 1.45 & 0.18 & 94.3 & 2.1\\
 10 &  J050754-610442 & J050754-610443 &  5.1318527 & -61.0785789 & . & 273.49 & 19.16 & 13.59 & 0.99 & 4.97 & 0.50 & 64.4 & 2.8\\
 11 &  J051637-723707 & - &   5.2772115 & -72.6188278 & . & 204.64 & 14.33 & 6.94 & 0.53 & 3.39 & 0.35 & 16.7 & 2.7 \\
 12 &  J051644-620706 & J051644-620702 &  5.2791331 & -62.1183586 & . & 653.56 & 45.79 & 24.77 & 1.76 & 3.79 & 0.38 & -24.1 & 2.8\\
 13 &  J052234-610757 & J052233-610800 &  5.3762222 & -61.1324997 & . & 146.58 & 10.27 & 0.97 & 0.10 & 0.66 & 0.08 & 29.2 & 3.4\\ 
 14 &  J053435-610606 & J053435-610605 &  5.5765971 & -61.1019211 & . & 170.48 & 11.94 & 5.47 & 0.49 & 3.21 & 0.37 & 35.2 & 2.4\\
 15 &  J054641-641522 & J054642-641513 &  5.7782806 & -64.2561417 & . & 20.36 & 1.43 & $<$0.18 & 0.06 & $<$0.91 & 0.30 & - & -\\
 16 &  J055009-573224 & J055009-573226 &  5.8359914 & -57.5401688 & . & 814.86 & 57.09 & 60.62 & 5.25 & 7.44 & 0.83 & 33.7 & 2.5\\
 17 &  J060755-603152 & J060755-603154 &  6.1320002 & -60.5311699 & . & 244.42 & 17.12 & 3.83 & 0.37 & 1.57 & 0.19 & -38.8 & 2.2\\
 18 &  J061030-605838 & J061030-605841 &  6.1750778 & -60.9773293 & . & 81.64 & 5.72 & 1.54 & 0.16 & 1.88 & 0.24 & - & -\\
 19 &  J062005-610732 & J062004-610737 &  6.3347946 & -61.1256409 & . & 120.89 & 8.47 & 12.65 & 0.92 & 10.46 & 1.05 & 70.4 & 2.8\\
 20 &  J062153-593509 & J062153-593510 &  6.3647527 & -59.5859718 & . & 89.93 & 6.30 & 0.43 & 0.08 & 0.48 & 0.09 & -20.2 & 5.4\\
 21 &  J062307-643620 & J062307-643624 &  6.3854808 & -64.6057205 & . & 285.89 & 20.03 & 10.81 & 1.02 & 3.78 & 0.44 & -7.2 & 2.2\\
 22 &  J062524-602030 & J062523-602025 &  6.4234222 & -60.3416901 & . & 81.30 & 5.70 & 0.23 & 0.07 & 0.28 & 0.09 & - & -\\
 23 &  J062857-624845 & J062857-624851 &  6.4826421 & -62.8125610 & . & 222.78 & 15.62 & 4.41 & 0.33 & 1.98 & 0.20 & -70.4 & 2.9\\
 24 &  J063546-751616 & J063547-751617 &  6.5962026 & -75.2713318 & pe & 1199.43 & 83.99 & 18.86 & 1.60 & 1.57 & 0.17 & -12.2 & 2.5\\
 25 &  J064428-671257 & J064428-671253 &  6.7411112 & -67.2160568 & . & 410.77 & 28.77 & 4.78 & 0.48 & 1.16 & 0.14 & -1.8 & 2.0\\
 26 &  J070031-661045 & J070031-661043 &  7.0086578 & -66.1791916 & . & 599.90 & 42.01 & 22.92 & 2.28 & 3.82 & 0.47 & 88.3 & 2.0\\
 27 &  J071509-682957 & J071511-683010 &  7.2526306 & -68.4993134 & . & 168.92 & 11.83 & 4.06 & 0.40 & 2.41 & 0.29 & 41.8 & 2.1\\
 28 &  J073856-673551 & J073856-673550 &  7.6489721 & -67.5975037 & . & 99.07 & 6.94 & 1.85 & 0.18 & 1.86 & 0.23 & 49.8 & 2.2\\
 29 &  J074331-672625 & J074332-672628 &  7.7254445 & -67.4404678 & pe & 198.09 & 13.88 & 10.26 & 1.03 & 5.18 & 0.63 & 1.0 & 2.0\\
 30 &  J074420-691906 & J074421-691908 &  7.7389582 & -69.3184662 & . & 155.15 & 10.87 & 15.22 & 1.22 & 9.81 & 1.04 & 59.6 & 2.6\\
 31 &  J075714-735308 & J075714-735306 &  7.9539001 & -73.8857498 & . & 58.83 & 4.12 & 2.00 & 0.20 & 3.40 & 0.42 & - & -\\
 32 &  J080633-711217 & J080632-711215 &  8.1094167 & -71.2047501 & pe & 84.07 & 5.89 & 0.33 & 0.05 & 0.40 & 0.06 & -66.9 & 4.5\\
\hline
\end{tabular}
\label{tab:catalogue}
}
\end{table*}

\section{Data analysis}
\label{sec:dataanaly}
We adopted a $3\sigma$ limit for detections in polarization. The median
sensitivity in polarization for our ALMA observations (including the
calibration error), is $\simeq 0.4\,$mJy. We achieved a detection rate of
$\simeq 97\%$: only $1$ object is non-detected, AT20GJ054641-641522. This is a quasar that went undetected in polarization also by our ATCA observations in both the 2014 and the 2016 campaigns, with $5\,\sigma$ detection limits in the $33-38\,$GHz band of $0.7\,$mJy and $2\,$mJy, respectively.

In the following sub-sections we discuss the polarimetric properties of our sample, combining observations from $2\,$GHz \citep[epoch: 2016 March and April,][]{Galluzzi2018}, through the $5.5-38\,$GHz range \citep[epochs: 2014 September, 2016 March and April, presented in][and new observations of 2016 July]{Galluzzi2017,Galluzzi2018} and up to $104.5\,$GHz (ALMA observations, 2016 August and September). In the analysis of total intensity spectra we include GLEAM data. We exclude from the analysis the FR-II source AT20GJ040848-750720, which was resolved by ALMA. ALMA observations of this source are presented in Section~\ref{sec:pecobj}.

\subsection{Spectral behaviour}
\label{subsec:spebeh}

The ATCA and ALMA observations are not simultaneous. While ALMA observations
were carried out at the end of August and at the end of September 2016, ATCA
observations at $33-38\,$GHz were performed at the beginning of April 2016 for
half of the present sample, and at mid July 2016 for the other half. The whole
sample of $32$ objects was observed at $2.1$ and $5.5-9\,$GHz in 2016 March-April, and only $13$ objects have measurements repeated in July.

At frequencies higher than $20\,$GHz, variability frequently exceeds $10\%$
even on time scales of few months. Therefore we have not attempted a joint fit
of ALMA and ATCA data, also on account of the $\sim 50\,$GHz frequency gap
between the two data sets.

Figure~\ref{fig:Spettri1} shows, for each source in our sample, a collection of
total intensity and polarization measurements. At the bottom of each panel, we also display a plot of the linear polarization fractions and, below each panel, a plot of the position angles
as a function of frequency. Together to ALMA data, we display measurements collected during ATCA 2014 and 2016 observations.
Moreover, in total intensity we include GLEAM \citep{HurleyWalker2017}, the Sydney University Molonglo Sky Survey \citep[SUMSS,][]{Mauch2003} and PACO \citep{Massardi2016} flux densities.

% The plotted data in total intensity include: the ALMA and ATCA 2016
% measurements (filled red circles); the GLEAM flux densities (filled orange
% circles); the Sydney University Molonglo Sky Survey (SUMSS) flux density
% (filled yellow circle); the ATCA 2014 observations (orange squares); the PACO
% observations (2009-2010; blue stars). 
% %\textbf{AT20G best epoch (2004-2008) observations? DA VERIFICARE}
% 
% The polarization data include:  ATCA 2016 and ALMA observations (filled black
% circles; upper limits are shown as filled downwards-pointing black triangles);
% ATCA September 2014 observations (black squares).
% %\textbf{AT20G best epoch (2004-2008) observations? DA VERIFICARE}
% 
% In the linear polarization fraction panels, filled purple squares refer to
% ALMA and ATCA 2016 observations and empty purple squares refer to September 2014 ATCA
% observations. Upper limits are shown as downward purple triangles.
% 
% As for the polarization angle, the filled indigo diamonds refer to ATCA 2016
% and ALMA observations and the indigo squares to the September 2014 ATCA
% observations.

The ALMA total intensity flux densities of most ($26$ out of $32$) sources are
somewhat in excess of expectations based on fits of the ATCA 2016 total
intensity measurements. The median excess is of $\sim 46\%$ (with a maximum of
$\simeq 98\%)$. The polarization fraction however indicates that we are still
dealing with synchrotron emission from the active nucleus. The unpolarized
free-free and the weakly polarized dust emission associated to star formation
in the host galaxies are expected to be much fainter. The excess is thus
suggestive of a different component coming out at a few mm wavelengths.

For $3$ sources (namely, AT20GJ035547-664533, AT20GJ053435-610606 and AT20GJ055009-573224) the absolute value of the flux density difference is
less than $10\%$, and may be accounted for by variability and/or measurement
errors. Again, only $3$ objects, namely AT20GJ050754-610442, AT20GJ051644-620706 and
AT20GJ062307-643620, have ALMA flux densities significantly fainter than expected. The
deficits are of $\simeq 10\%$, $18\%$ and $81\%$, respectively. However, even
in the latter case there is no sign of a spectral break and the median spectral index in total intensity
between $36.5$ and $97.5\,$GHz is $\alpha_{36.5}^{97.5}\simeq -0.19$ (we use
the convention $S_\nu \propto \nu^\alpha$). This result is in very good agreement with the predictions of the C2Co model \citep[][see their Table 6]{Tucci2011},
albeit in a slightly different frequency interval, namely 30-100 GHz.%\footnote{This outcome is also in remarkable agreement with the recent findings presented in \citet[][Int. LV; Section 5.2.1 and Figure 4]{PlanckIntermLV}, that point towards a better fit of the C2Co model of Tucci et al. (2011) to the most recent data in differential number counts, obtained by applying the ``Matrix Filters'' detection method to the full mission all-sky temperature maps produced by the ESA {\it Planck} satellite mission.}

%\clearpage
%-----------------------------Figure Start---------------------------
%\begin{landscape}
\begin{figure*}
%\vspace{-1.0cm}
\centering
\includegraphics[scale=0.85]{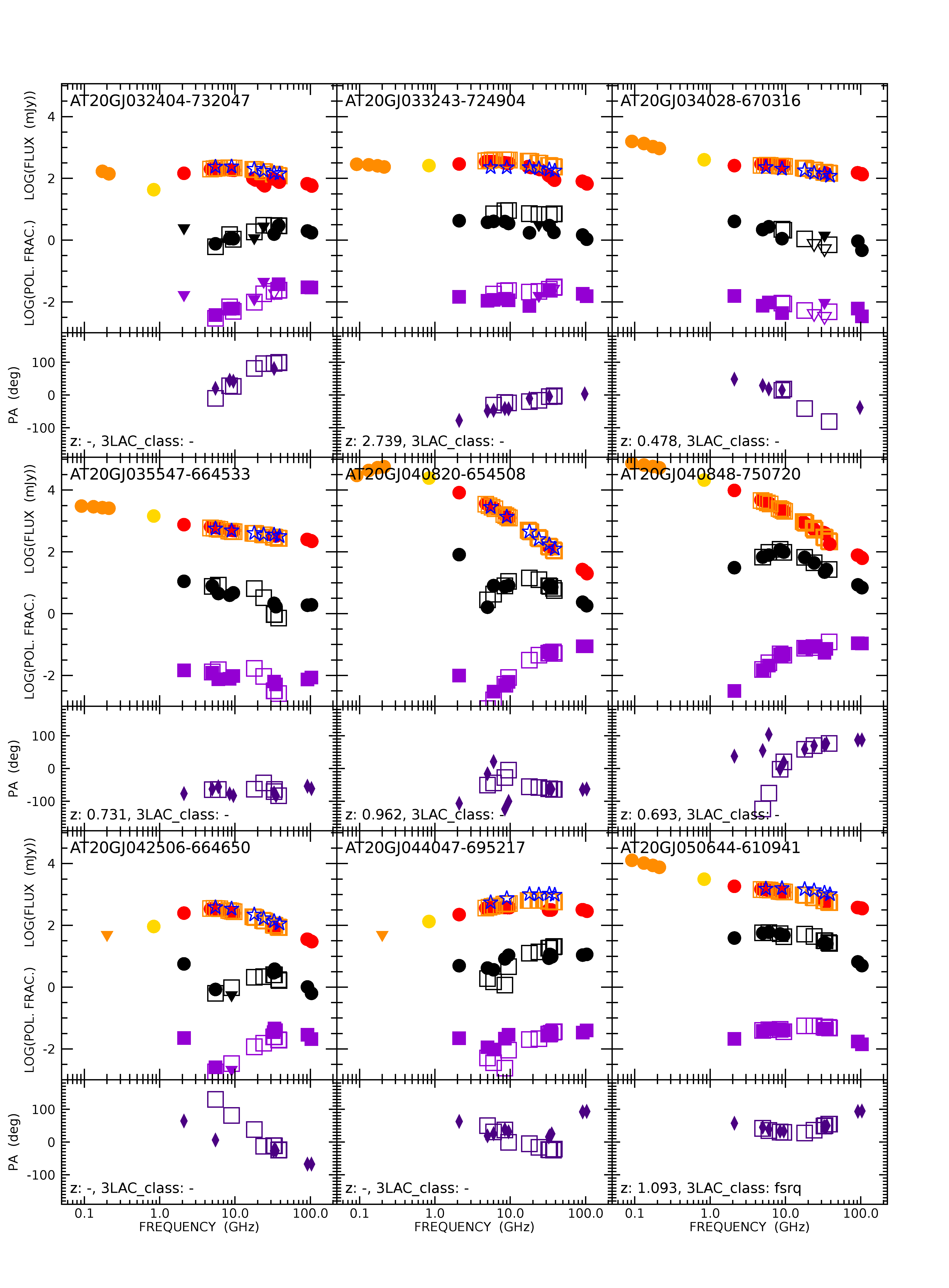}
%\vspace{-0.85cm}
\caption{Spectra in total intensity, in polarization, in polarization fraction and in
polarization angle for the $32$ objects of the faint PACO sample, observed with ALMA. The error bars are not displayed since they are smaller than the symbols. {\bf Total intensity:} filled red circles indicate ATCA 2016 and ALMA observations. The filled orange and yellow circles show GLEAM and SUMSS flux densities, respectively. The orange squares are ATCA 2014 observations and the blue stars are PACO observations (2009-2010). {\it (Continued...)}}
\label{fig:Spettri1}
\end{figure*}
%\end{landscape}
%-----------------------------Figure End------------------------------
\addtocounter{figure}{-1}
%-----------------------------Figure Start---------------------------
%\begin{landscape}
\begin{figure*}
%\vspace{-2.5cm}
\centering
\includegraphics[scale=0.85]{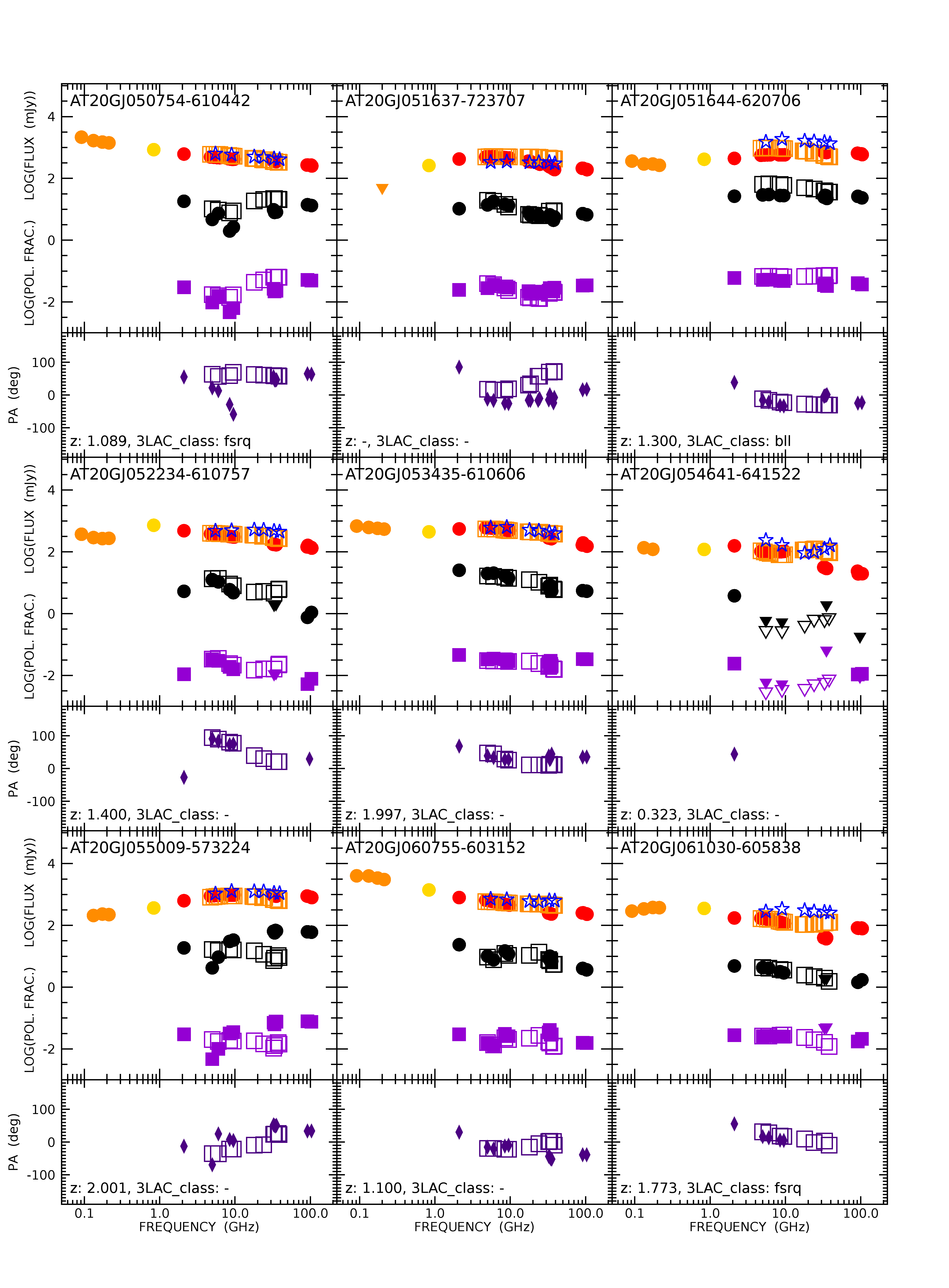}
\caption{{\it (Continued.)} {\bf Polarization (flux density):} filled black circles
refer to ATCA 2016 and ALMA observations. Upper limits are shown as black filled
downwards-pointing triangles. Black squares represent previous ATCA observations (September 2014). Their upper limits are displayed as downwards-pointing empty black triangles. {\bf Linear polarization fractions:} filled purple squares, with upper limits shown as downwards-pointing filled triangles, for ALMA and ATCA
observations; purple squares with upper limits shown as downwards-pointing empty purple triangles
for September 2014 ATCA observations. {\it (Continued...)}}
\label{fig:Spettri2}
\end{figure*}
%\end{landscape}
%-----------------------------Figure End------------------------------
\addtocounter{figure}{-1}
%-----------------------------Figure Start---------------------------
%\begin{landscape}
\begin{figure*}
%\vspace{-2.5cm}
\centering
\includegraphics[scale=0.85]{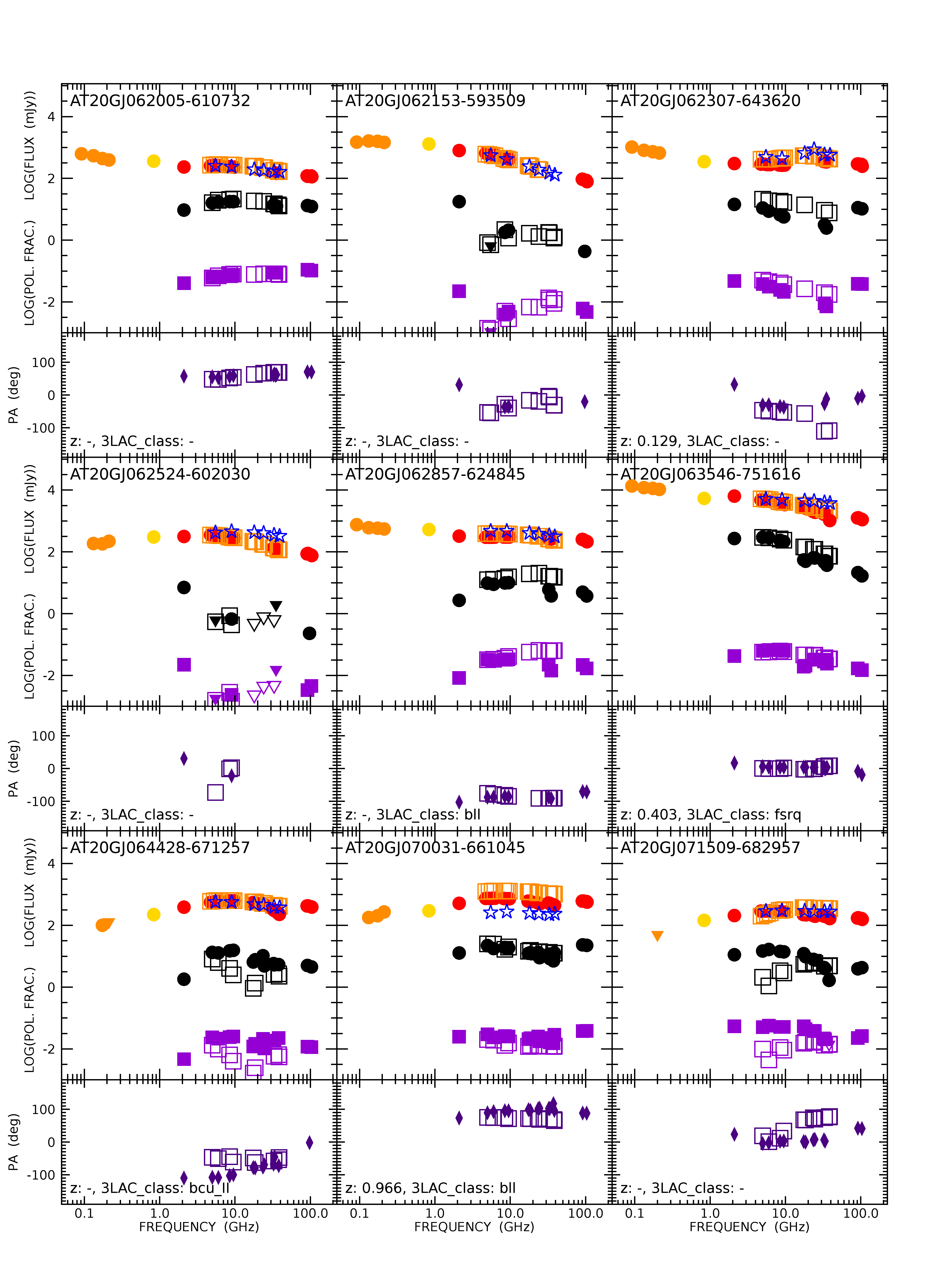}
\caption{{\it (Continued.)} {\bf Polarization angle:} filled indigo diamonds
for ATCA 2016 and ALMA observations; indigo squares for September 2014 ATCA observations. For each object (where available) we also report the redshift z and the classification provided by the Third Catalog of active galactic nuclei (3LAC) released by the Fermi Collaboration \citep{Ackermann2015}.}
\label{fig:Spettri3}
\end{figure*}
%\end{landscape}
%-----------------------------Figure End------------------------------
\addtocounter{figure}{-1}
%-----------------------------Figure Start---------------------------
%\begin{landscape}
\begin{figure*}
%\vspace{-2.5cm}
\centering
\includegraphics[scale=0.85]{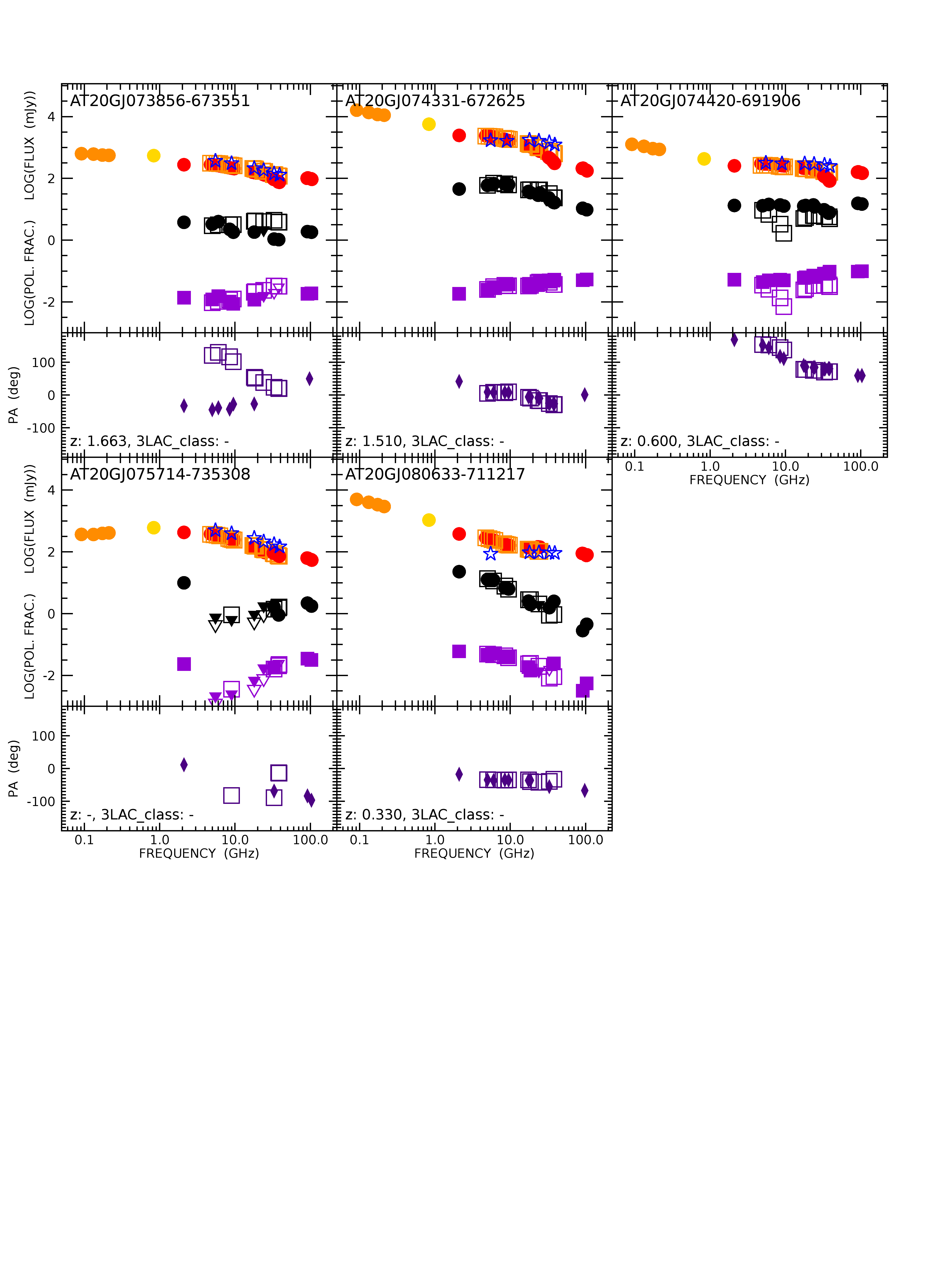}
\caption{{\it (Continued.)}}
\label{fig:Spettri4}
\end{figure*}
%\end{landscape}
%-----------------------------Figure End------------------------------

\begin{figure}
%\vspace{-2.5cm}
\centering
%\includegraphics[width=\columnwidth,trim={0 0 2.5cm 2.0cm},clip]{CCIP36100_zoomed_resubmission.eps}
%\vspace{-0.85cm}
\includegraphics[width=\columnwidth,trim={0 0 0 0},clip]{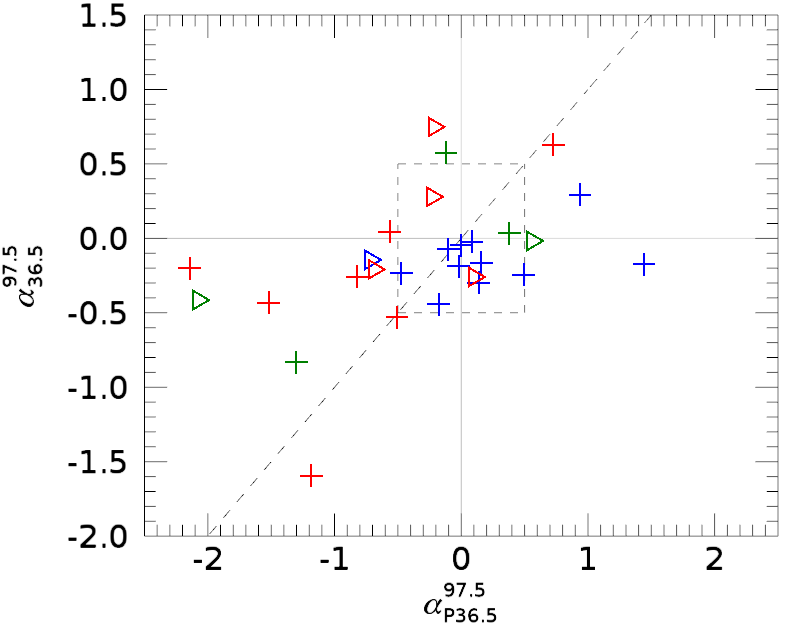}
\caption{Colour-colour plot showing spectral indices in total intensity versus those in polarization.
Different colours refers to different sub-classes in total intensity: red for steep-,
green for peaked- and blue for flat-spectrum objects \citep[\textit{cf.} the classification introduced by][]{Galluzzi2018}. Rightward-pointing triangles are for lower limits for spectral indices in polarization (non-detections at ATCA frequencies). The central square delimited by dashed lines represents the region in which both spectral indices are between $-0.5$ and $0.5$. The other dashed line is a bisector.}
\label{fig:CCIP36100}
\end{figure}

%\FloatBarrier
Figure~\ref{fig:CCIP36100} compares the spectral indices in total intensity and
in polarization between 36.5\,GHz (the central frequency of ATCA 2016
observations) and 97.5\,GHz (the central frequency of ALMA observations). Total
intensity spectral indices, $\alpha_{36.5}^{97.5}$, are, with a few exceptions,
in the range $-0.50$--$0.50$. In polarization there are a couple of sources with spectral indices, $\alpha_{{\rm p},36.5}^{97.5}$ as steep as $-1.5$ or even $-2.0$.

There are also two sources with $\alpha_{{\rm p},36.5}^{97.5}\ge 1$ and $7$
sources undetected in polarization at $35$ or $38\,$GHz but detected at
$97.5\,$GHz, i.e. with only a lower limit to $\alpha_{{\rm p},36.5}^{97.5}$.
Only part of these lower limits may be understood in terms of the higher
sensitivity of ALMA observations compared to the ATCA ones. In other cases they
provide further support to indications of an additional synchrotron component
showing up at frequencies of $\sim 200-300\,$GHz in the source frame.
%To fit the spectra we used the same functional forms (a double power-law and a triple power-law) of \citet{Galluzzi2017}, and adopted similar criteria about the minimum number of observations to properly constrain the fit parameters. Given the small fraction (less than $10\%$) of non-detections we did not consider the upper limits in doing the spectral fits. About total intensity, $81\%$ of the spectra could be successfully fitted in this way. In polarization we have two cases (AT20GJ054641-641522, AT20GJ062524-602030) in which we do not have enough detections in polarization to run the fitting procedures. The success rate results (as already found) lower with respect to total intensity, i.e. $\simeq 53\%$.

%Similarly to what found for the enlarged sample, considering all the spectra (both in total intensity and polarization) most ($\simeq 52\%$) of our source spectra could be fitted with a double power-law down-turning at high frequencies. An upturning double power-law was required in $4$ cases and a triple power-law in $16$ cases. The median values of the reduced $\chi^2$ are $1.5$ and $4.2$ for $I$ and $P$, respectively.

%-----------------------------Figure Start---------------------------
%\begin{landscape}
%\end{landscape}
%-----------------------------Figure End------------------------------
\subsection{Linear polarization fraction}
\label{subsec:ALMApolfrac}

%The knot-like synchrotron emission at higher frequencies/energies are generally
%closer and closer to the base of the AGN jet. In those regions the ejection
%speed are typically closer to the speed of light and magnetic field should be
%more ordered to support the radiative process. As a consequence, an increase of
%the polarization fraction is, at least in principle, expected with frequency in
%these regimes. There is indeed another phenomenon which is poorly understood
%even in the non-relativistic regime: the turbulence. Its typical effect
%consists in an energy transport from higher spatial scales towards dissipation
%on smaller scales and, depending on the particular spectrum assumed, it may
%hamper ordering effects in magnetic field, lowering the polarization fraction.
%Any structural inhomogeneity (in the emitting plasma and/or magnetic field
%configuration) within the synthesized beam lowers the polarization fraction as
%well. Moreover, the high RMs found by \citep{Pasetto2016,Galluzzi2018} at
%frequencies $> 20\,$GHz seem to suggest that non-relativistic electrons in the
%narrow-line region clouds might act as efficient Faraday screens for emitting
%knots closer to the jet basis. However, as stressed in the previous chapters,
%the lack of polarimetric data on large complete samples prevented any firm
%conclusion about the existence or not of such trend. Here we provide one of the
%first unbiased assessment about observed polarization fractions at $\sim
%100\,$GHz taking into account our complete sample of $32$ objects.

The median polarization fraction measured by ALMA for the full sample is
$2.2\pm 0.6\%$, close to the median value at $38\,$GHz ($2.09\%$) for the
larger sample of $104$ objects \citep[\textit{cf.}][]{Galluzzi2018}.
Our result is in good agreement with estimates based on \textit{Planck} maps at $100\,$GHz obtained by applying stacking techniques by \citet[][$1.8\,(+0.4,-0.3)\%$]{Bonavera2017} and by
using intensity distribution analysis (IDA) method by \citet[][$1.8\pm 0.5$\%]{Trombetti2018}.

\begin{figure}
%\vspace{-2.5cm}
\centering
\includegraphics[width=0.9\columnwidth]{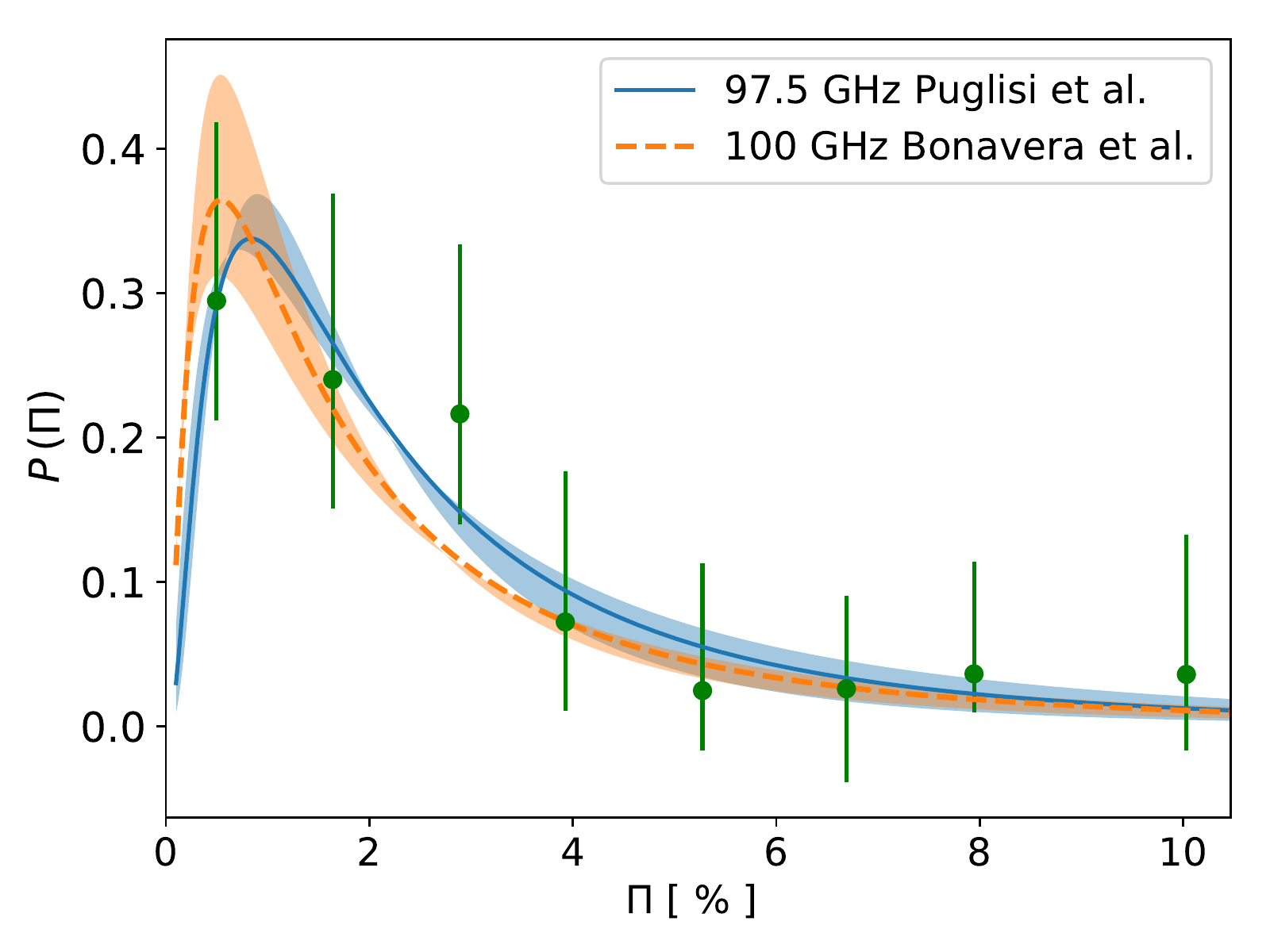}
\vspace{-0.5cm}
\caption{Distribution of the percentage polarization fraction at $97.5\,$GHz obtained with
a bootstrap and re-sampling of the observed distribution of the polarization fractions (green points).
The lognormal fit is shown by the blue solid line \citep[the shaded area represents the $1\sigma$ uncertainty in the fitting curve, already presented by][]{Puglisi2018}. We also plot the distribution obtained by \citet[][the orange dashed line with the corresponding $1\sigma$ shaded area]{Bonavera2017}.}
\label{fig:BotResPPol100}
\end{figure}

We also estimated the distribution of the percentage polarization fraction,
$\Pi$, using a bootstrap and re-sampling approach. Each detection was
associated with the mean value of a Gaussian with $\sigma$ given by the error
on the polarization fraction. When only an upper limit is available, we used a
uniform distribution between $0$ and the $3\sigma$ upper limit. Then, we generated $1000$ simulated data sets by resampling with repetitions the
distributions of percentage polarization fractions of each source. The results
of the simulation are reported in Fig.~\ref{fig:BotResPPol100} and in
Table~\ref{tab:ObsPPol100}. In Fig.~\ref{fig:BotResPPol100} we also show the
best fit log-normal function:
\begin{equation}\label{eq:lognorm}
\mathcal{P}(\Pi)=\frac{A}{\Pi\sigma\sqrt{2\pi}}\exp{\left[-\frac{\ln^2(\Pi/\mu)}{2\sigma^2}\right]},
\end{equation}
with $A=0.86$, $\mu=2.05$ and $\sigma=0.97$.

%-----------------------------Figure Start---------------------------
%\begin{landscape}
%\end{landscape}
%-----------------------------Figure End------------------------------

\begin{table}
\caption{Distribution of the percentage polarization fractions at $97.5\,$GHz outcoming from the bootstrap and re-sampling performed on the ALMA data.}
\label{tab:ObsPPol100}
\centering
\begin{tabular}{cccc}
\hline
$\Pi$ (per cent)& Probability & lower & upper\\
& & error & error\\
\hline
0.502 & 0.299 & 0.083 & 0.123 \\
1.634 & 0.235 & 0.089 & 0.129 \\
2.897 & 0.216 & 0.077 & 0.118 \\
3.943 & 0.074 & 0.062 & 0.104 \\
5.269 & 0.026 & 0.042 & 0.088 \\
6.667 & 0.024 & 0.064 & 0.065 \\
7.960 & 0.036 & 0.027 & 0.078 \\
10.044 & 0.037 & 0.053 & 0.097 \\
\hline
% 0.502067404464 & 0.299032258065 & 0.0831113706091 & 0.123497649185 \\
% 1.63444752297 & 0.235161290323 & 0.0891458488771 & 0.129032258065 \\
% 2.89680182494 & 0.215677419355 & 0.0766053654361 & 0.117604881002 \\
% 3.94270911206 & 0.073935483871 & 0.0615818177795 & 0.104389289597 \\
% 5.2690030508 & 0.0255483870968 & 0.0415476017534 & 0.0881306712119 \\
% 6.66681315539 & 0.0237096774194 & 0.0645161290323 & 0.0645161290323 \\
% 7.96040452527 & 0.0359677419355 & 0.0267268463206 & 0.0778778568507 \\
% 10.0443758644 & 0.0372903225806 & 0.0525052398025 & 0.0967741935484 \\
\end{tabular}
\end{table}

%\begin{figure}
%%\includegraphics[width=\columnwidth]{PPolAllDATA-eps-converted-to.pdf}
%%\includegraphics[width=\columnwidth]{PPolSteepDATA-eps-converted-to.pdf}
%%\includegraphics[width=\columnwidth]{PPolPeakedDATA-eps-converted-to.pdf}
%%\includegraphics[width=\columnwidth, trim={0cm 1.5cm 3.5cm 3.5cm},clip]{PPolASPF.eps}
%%\includegraphics[width=1.0\columnwidth]{PPolASPF_w100.eps}
%\includegraphics[width=\columnwidth]{PPolASPF_w100.jpg}
%\caption{Frequency dependence of the median polarization fraction for sources with different spectral classification. At
%$2.1$, $5.5$, $9$, $18$, $24$, $33$ and $38\,$\,GHz we show the median polarization percentages,
%with their uncertainties, for the full sample by \citet{Galluzzi2018}, comprising a total of $104$ sources.
%To these estimates we have added the median polarization percentage at $97.5\,$GHz
%for the complete sub-sample of $32$ objects observed with ALMA.  }.
%\label{fig:PPolVsfreq2}
%\end{figure}

\begin{figure}
\includegraphics[width=\columnwidth]{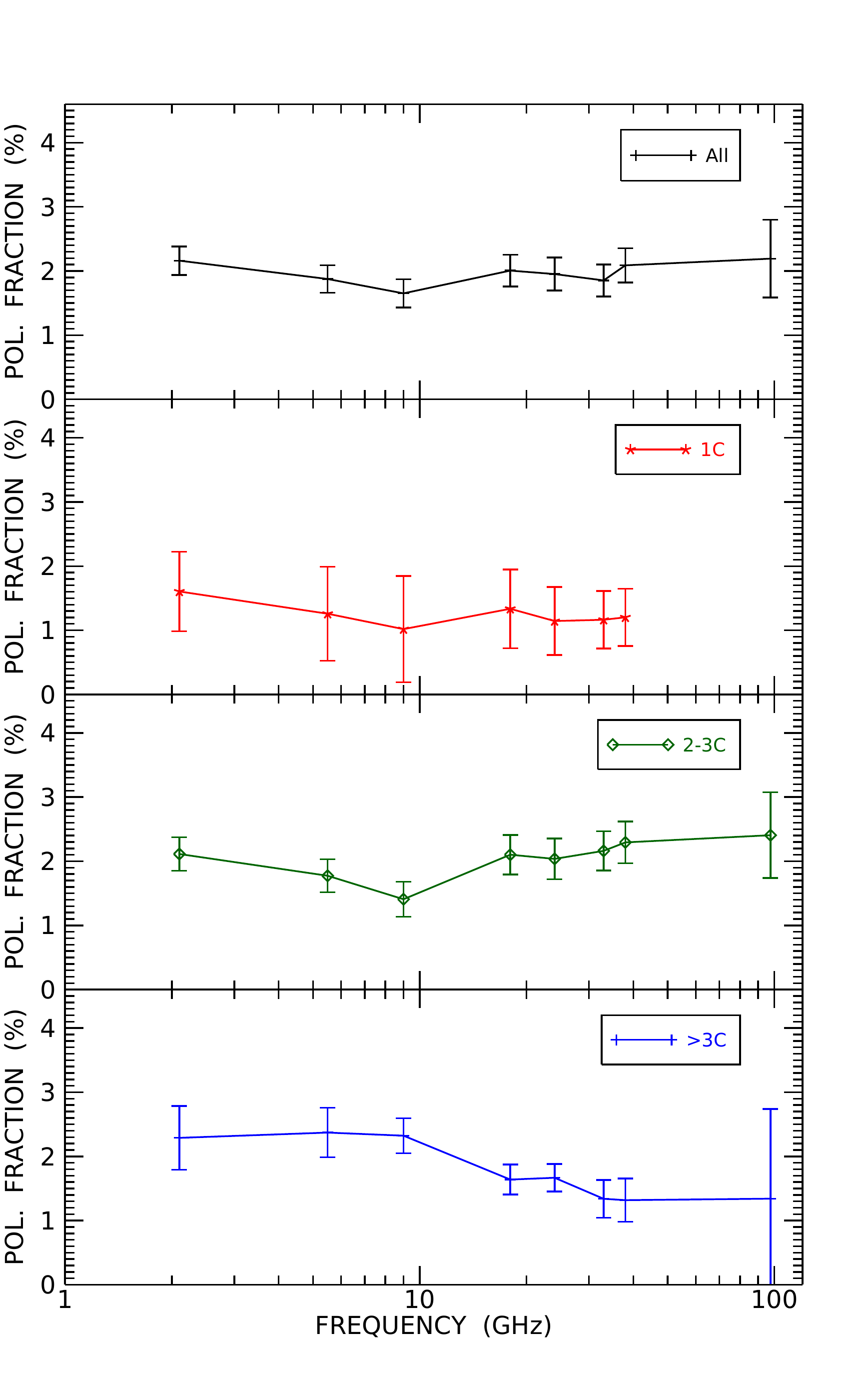}
\vspace{-0.75cm}
\caption{Frequency dependence of the median polarization fraction for sources with different spectral classification. At
$2.1$, $5.5$, $9$, $18$, $24$, $33$ and $38\,$\,GHz we show the median polarization percentages,
with their uncertainties, for the larger sample studied by \citet{Galluzzi2018}, comprising a total of $104$ sources. To these estimates we have added the median polarization percentage at $97.5\,$GHz
for the complete sub-sample of $32$ objects observed with ALMA. We also subdivide the sources by the number of spectral components: 1 (1C), 2--3 (2--3C) and more ($>\,$3C).}
\label{fig:PPolNCVsfreq2}
\end{figure}

In \citet{Galluzzi2018} we briefly discussed the 
%topic of spectral classification in this new era of a wider range of frequency coverage both in total intensity and polarization, arguing that purely spectral index based classifications are poor in investigating the physics of these objects. We thus
spectral classification in the presence of a wide frequency coverage both in total intensity and polarization. For every object (in our sample of $104$ compact extragalactic sources) we compared the spectrum in total intensity with that in linear polarization, finding in more than $90\%$ of cases signs of $2-3$ synchrotron components (e.g. multiple bumps in the spectra or features appearing only in polarization, where total intensity still looks smooth). Thus, we classified objects which can be explained in terms of a single emitting region as ``1C'', those with $2$ or $3$ components as ``2-3C'', and sources with indications of more than 3 components (typically in the range $70\,$MHz -- $\sim 30\,$GHz)  as ``$>\,$3C''.
 
Here we complement the analysis about the frequency dependence of the median polarization fraction provided by \citet{Galluzzi2018} by investigating this aspect at higher frequencies (i.e. $97.5\,$GHz). We again apply the same classification in terms of synchrotron components in order to distinguish between sub-populations. However, we warn the reader that this classification is based on polarimetric data
collected in the 2014 campaign of ATCA observations. We were not able to update this because of the lack of $18-24\,$GHz polarimetric observations for several objects in 2016 campaigns and because ALMA observations are not strictly co-eval to 2016 ATCA ones (variability might bias the classification). Our sample of $32$ objects displays $27$ sources classified as 2-3C and $5$ sources in the $>\,$3C class. We found no 1C objects.
%From the other plot for the polarization fraction behaviour in frequency , the other classification criterion provided in \citet{Galluzzi2018}, i.e. based on spectral indices $\alpha_{2.5}^{5.5}$ and $\alpha_{28}^{38}$ indeed reports a spectral change between the 2014 and 2016 epochs for several sources which might bias the comparison with polarization fractions presented at ATCA frequencies (i.e. up to $38\,$GHz) .   
% the spectral classification provided by \citet{Galluzzi2018}, based on the ATCA observations for these objects. Sources were classified according to their 2.5 to 5.5\,GHz
%and 28 to 38\,GHz spectral indices in five spectral types. Only 3 types (flat-,
%steep- and peaked-spectrum) are represented in our sample. Note that our
%steep-spectrum sources are compact, at variance with the classical
%steep-spectrum sources, generally classified on the basis of the 1.4 to
%4.8\,GHz spectral index, that are normally extended.
The results are presented in Fig.~\ref{fig:PPolNCVsfreq2}. The median
polarization percentages at $2.1$, $5.5$, $9$, $18$, $24$, $33$ and $38\,$\,GHz
refer to the larger sample analyzed by \citet{Galluzzi2018}, comprising a total of $104$
sources. The median polarization percentage at $97.5\,$GHz is for the ALMA sample of $32$ objects. The errors on median values are given by $1.253\, {\rm rms}/\sqrt{N}$, where rms is the standard deviation
around the mean and $N$ is the number of objects \citep[\textit{cf.}][]{Arkin1970}. The
error bars at $97.5\,$GHz are larger since the size of the sample is smaller by
a factor $\sim 3$ with respect to lower frequencies.

As illustrated by the Fig.~\ref{fig:PPolNCVsfreq2}, the data do not indicate any statistically
significant trend with frequency for all the objects. According to the analyses
by \citet{Bonavera2017} and \citet{Trombetti2018}, the median polarization
fraction remains essentially frequency independent over the full range of
\textit{Planck} polarization measurements (30--353\,GHz). Moreover, negligible frequency dependency has been found by \citet{Puglisi2018} by combining data in a wide range of frequencies (from $1.4$ to $217\,$GHz).
%There are weak indications of slightly different spectra of the polarization fraction among different sub-populations. In particular, the polarization fraction is somewhat higher at 97.5\,GHz than a tens of GHz. SIGNIFICATIVITA' DELLA DIFFERENZA?

%We have also looked for possible differences in the frequency dependence of the polarization fraction among sources sub-classes based on the number of synchrotron components we inferred from total intensity and polarization spectra. In this case the classification is based on spectro-polarimetric data collected in the 2014 campaign of ATCA observations.

As pointed out by \citet{Galluzzi2018}, sources with 2--3 spectral components
(2--3C) seem to show a minimum of the polarization fraction at $\sim 10\,$GHz
while for sources with more than 3 components ($>\,$3C) a slight decrease above
this frequency is indicated by the data. The ALMA measurements are consistent
(although with large uncertainties) with frequency independent polarization
fractions above some tens of GHz.

\begin{figure}
%\vspace{-2.5cm}
\centering
\includegraphics[width=1.0\columnwidth]{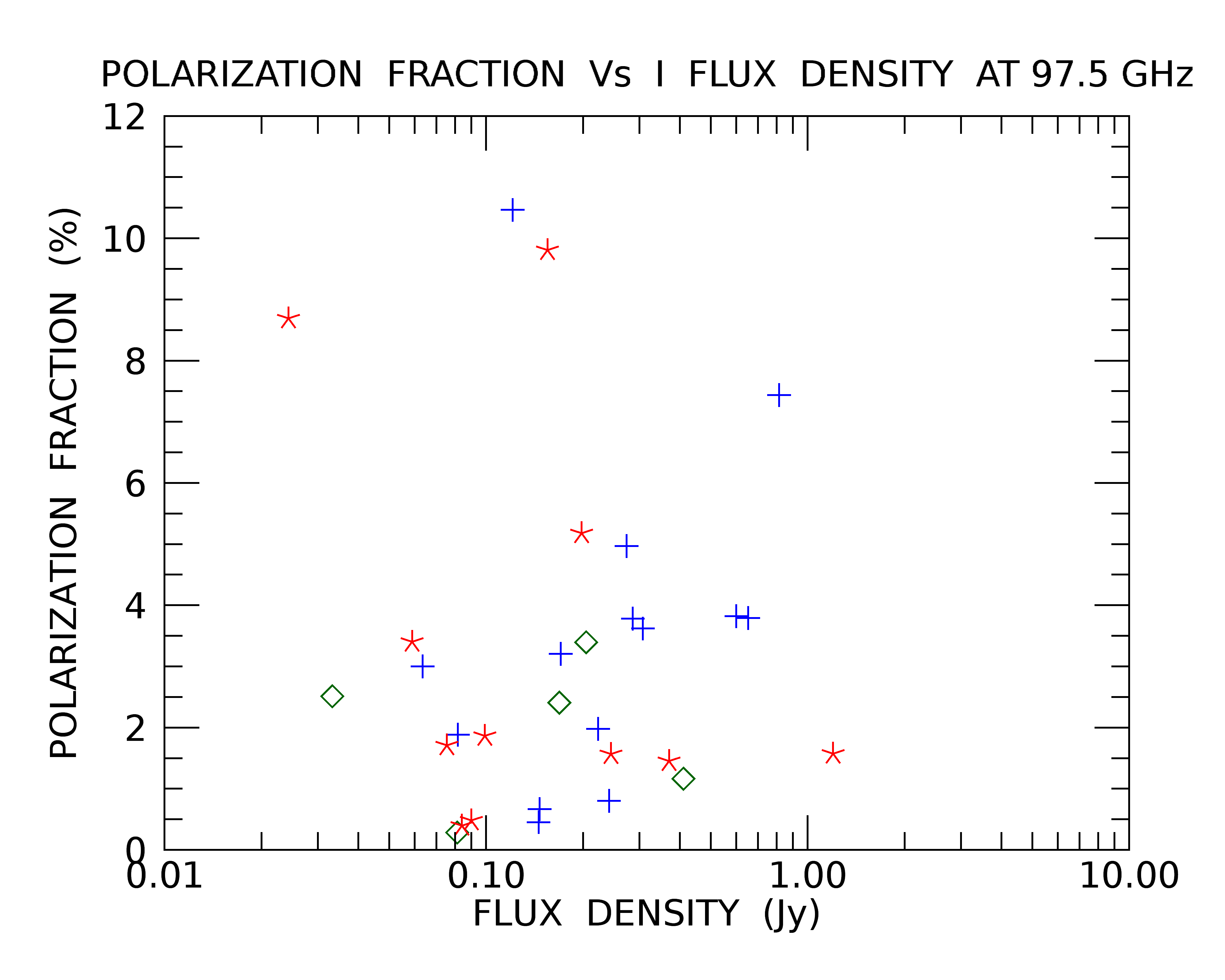}
\vspace{-0.5cm}
\caption{Polarization fraction against total flux density at $97.5\,$GHz for
the complete sample observed with ALMA. Red stars, green diamonds and blue pluses stand for steep-, peaked- and flat-spectrum objects respectively \citep[\textit{cf.}][]{Galluzzi2018}.}
\label{fig:PPolVsfluI100}
\end{figure}

\citet{Trombetti2018} also found no evidence of a dependence of the median
polarization fraction on the total flux density. As shown by
Fig.~\ref{fig:PPolVsfluI100} the ALMA data are consistent with this result:
there is no sign of a correlation between the polarization fraction and the
total flux density, neither for the full sample nor for steep-, peaked- and
flat-spectrum objects (identified by red stars, green diamonds and blue
pluses, respectively) separately. However, the small size of the sample
prevents any firm conclusion.
%-----------------------------Figure Start---------------------------
%\begin{landscape}
%\end{landscape}
%-----------------------------Figure End------------------------------
\subsection{Rotation measures at ALMA frequencies}
The sensitivity of our ALMA observations has allowed several detections in
Stokes $Q$ and $U$ with signal to noise ratios up to $\sim 10$ combining the
four $2\,$GHz bands. For $22$ objects out of $32$ both $Q$ and $U$ were detected
at a $\ge 6\,\sigma$ level, which in principle might allow us to have a
$3\,\sigma$ detection in each band. Three well determined polarization angles
are the minimum requirement to study the rotation measures (RMs) of our
sources. We have also attempted to split each band into two $1\,$GHz sub-bands,
bringing to 8 the maximum number of spectral measurements per source.

In the case of a foreground screen of magnetized plasma the polarization angle
varies as $\Delta \phi = RM\lambda^2$. The RMs were estimated using this
relation. 

\begin{table*}
\caption{\small List of the $11$ objects with an observed RM in ALMA Band 3 ($90-105\,$GHz) non-compatible with a null rotation at a $1\,\sigma$ level. We also report for each source the corresponding RMs found at lower frequencies (if any), the inferred number of synchrotron components  and the redshift, when available \citep[provided by][]{Galluzzi2018}. RMs and associated errors are in $\hbox{rad}\,\hbox{m}^{-2}$.}
\label{tab:ALMARMdetlist}
\small
\begin{tabular}{|l|lcr|c|c|c|}
\hline
(AT20G) name & & $RM_{\rm obs}$& & $\sigma_{\rm RM}$ & \# Comp & z\\
& $2-9$ & $18-38$ & $90-105$ & $90-105\,$GHz& &\\
\hline
J035547-664533 & -12 & - & 52075 & 13673 & $>$3 & 0.73\\
%J040820-654508 & -  & 600 & -11955 & 9849 & 2 & 0.96\\                         
J044047-695217 & - & 1500 & -6243 & 2550 & 2 & -\\
J050644-610941 & - & - & -9305 & 7992 & 2 & 1.09\\
J050754-610442 & - & 400 & 11593 & 10427 & 2 & 1.09\\ 
J051637-723707 & -21 & -3200 & -12039 & 10037 & 2 & - \\
J051644-620706 & 54 & 200 & -11976 & 10482 & 3 & 1.30\\
J053435-610606 & - & 0 & -14498 & 8696 & $>$3 & 2.00\\
j062307-643620 & 78 & - & -44998 & 5792 & 2 & 0.13\\
J063546-751616 & 16 & -800 & 85187 & 7860 & 2 & 0.40\\
%J070031-661045 & 0 & 200 & 1424 & 1184 & 3 & 0.97\\
J071509-682957 & - & 700 & 7131 & 5381 & 2 & -\\
J075714-735308 & - & - & 98273 & 10877 & 2 & - \\
\hline
\end{tabular}
\end{table*}

\begin{figure*}
%\vspace{-2.5cm}
\centering
\includegraphics[scale=0.90]{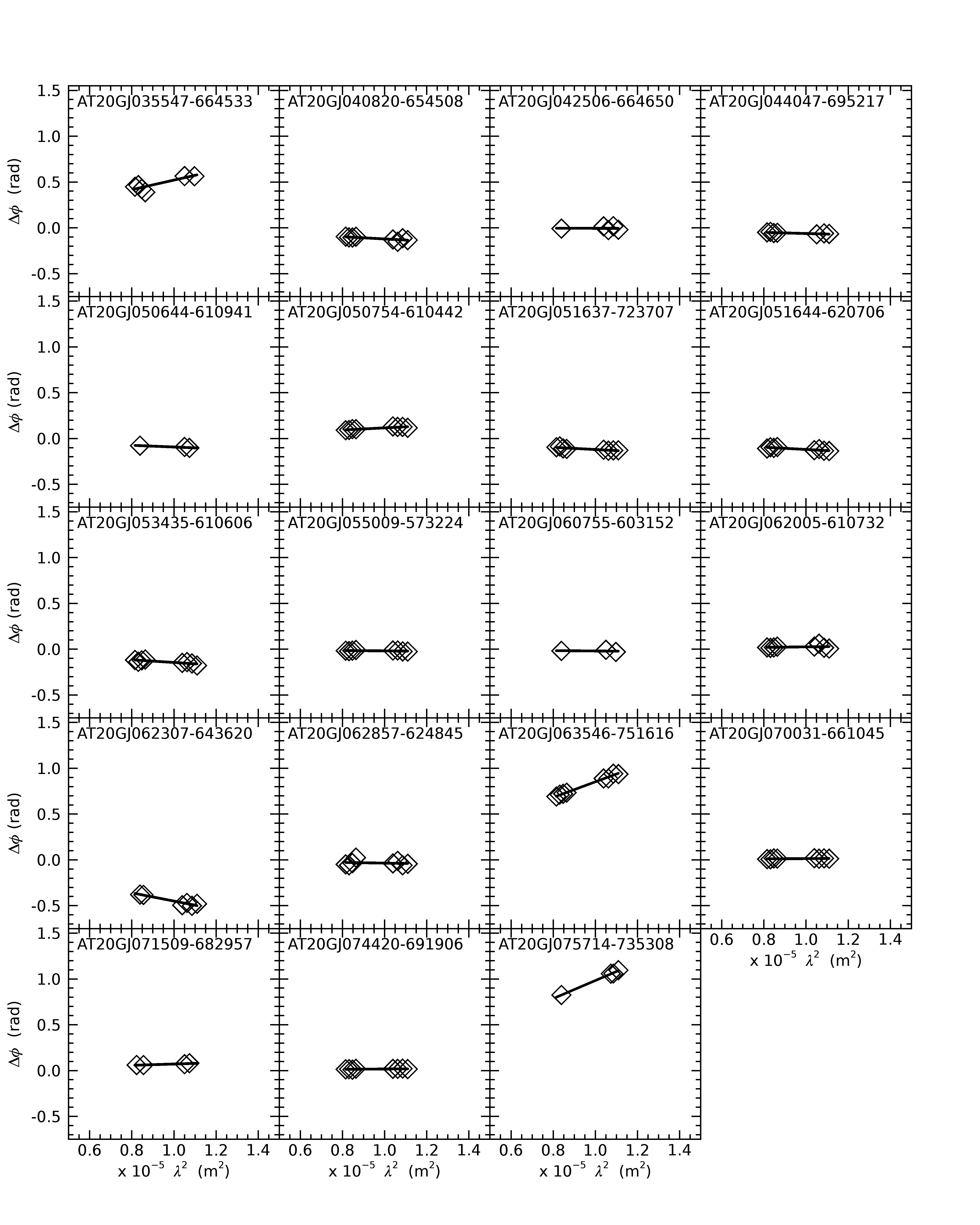}
\vspace{-0.5cm}
\caption{Successful RM fits for $19$ objects of the complete sample observed with ALMA between $90$ and $105\,$GHz.}
\label{fig:ALMARMfit}
\end{figure*}

Following \citet{Galluzzi2018} we used the IDL ``linfit'' procedure,
accepting only fits with a reduced $\chi^2<2$ and with a probability level $>
0.1$. In Fig.~\ref{fig:ALMARMfit} we show the $19$ successful fits. As discussed in \citet{Galluzzi2018} the $1/\lambda^2$ contribution to the uncertainty makes RM measurements extremely difficult at high frequencies. In our case only $11$ objects have RMs not compatible with $0$, at a $\ge 1\,\sigma$ level. In the Table~\ref{tab:ALMARMdetlist} we report the list of the observed RMs with the associated errors provided  by the fitting procedures. The median relative error for these cases is $\sim 60\%$ but we warn the reader that in four cases (i.e. AT20GJ050644-610941, AT20GJ050754-610442, AT20GJ051637-723707 and AT20GJ051644-620706) relative errors on RMs are as high as $80-90\%$.

%-----------------------------Figure Start---------------------------
%\begin{landscape}
%\end{landscape}
%-----------------------------Figure End------------------------------

In the upper part of Table~\ref{tab:ALMARM} we report the median values of the non-zero RMs derived from the above equation for these $11$ objects. For the $7$ objects with measured redshift we have computed the RMs at the source, correcting for the effect of redshift and for the relatively small contributions of our own Galaxy and of Earth's ionosphere, as detailed by \citet{Galluzzi2018}; the results are given in the lower part of the table. Also shown in the table are the results for the 2--3C and the
$>\,$3C sources considered separately (there are no 1C objects in the ALMA sample).

Although the number of objects is too small to reach any firm conclusion, we
note that the median RM  at the source ($\simeq
6.4\times 10^4\,\hbox{rad}\,\hbox{m}^{-2}$) is one order of magnitude higher than
that obtained for the $18-38\,$GHz frequency range and two orders of magnitude
higher than that found for the $2-9\,$GHz range \citep[\textit{cf.}][their Table
4]{Galluzzi2018}.

Our results seem to be still consistent with the indication of an increase of the
median RM with increasing number of spectral components, reported by
\citet{Galluzzi2018}. If confirmed, the extreme values derived from ALMA
measurements would require very dense screens of magnetized plasma \citep[\textit{cf.}][]{Hovatta2019}. Such
screens may heavily depolarize the radiation emitted at the basis of the
relativistic jet and thus offer an explanation for the lack of an observed
increase of the polarization fraction with increasing frequency. In fact, the
emission at higher and higher frequency is expected to come from regions closer
and closer to the nucleus where the magnetic field should be more ordered and
the polarization fraction correspondingly higher.

\begin{table}
\caption{Median values of the RMs between $90$ and $105\,$GHz. The upper part
of the table refers to the observed RMs for the $11$ sources with a non-null value at $1\,\sigma$ level. The lower part
gives the RMs at the source for the subset of objects for which redshift measurements
are available. In parenthesis are the numbers of objects in each group. RMs are in $\hbox{rad}\,\hbox{m}^{-2}$.}
\label{tab:ALMARM}
\centering
\begin{tabular}{|c|c|c|}
\hline
All sample ($11$)& 2-3C ($9$)& $>$3C ($2$)\\
\hline
$1.2\times 10^4$ & $1.2\times 10^4$ & $3.3\times 10^4$\\
\hline
All sample ($7$)& 2-3C ($5$)& $>$3C ($2$)\\
\hline
$6.4\times 10^4$ & $5.7\times 10^4$ & $1.4\times 10^5$\\
\hline
\end{tabular}
\end{table}

\section{Source Counts}\label{sec:soucou100ghz}

We have exploited our ALMA polarization measurements to derive the differential
source counts in polarization at $95\,$GHz, $n(P)\equiv dN/dP$. We started
from the C2Ex model for  total intensity source counts, $n(S)$, by
\citet{Tucci2011} and used the approach of \citet{Tucci2012}:
\begin{equation}
n(P)=\int_{S_0=P}^\infty {\mathcal{P}\left(m=\frac{P}{S}\right)n(S)\frac{dS}{S}},
\label{equ:difsoucouP}
\end{equation}
where $\mathcal{P}$ is the probability density distribution for the
polarization fraction $m=\Pi/100$, given by eq.~(\ref{eq:lognorm}). The
integration over $S$ is truncated at $S_0=P$, where the polarization fraction
is $100\%$; however the result is insensitive to the choice of $S_0$ (provided
that it is not much larger than $P$), since eq.~(\ref{eq:lognorm}) goes rapidly
to zero for $\Pi
> 10\%$.

\begin{figure}
\centering
%\includegraphics[width=0.8\columnwidth]{polariz_numbercounts_20ghz.pdf}
%\vspace{-0.85cm}
%\includegraphics[width=1.0\columnwidth]{polariz_numbercounts_95ghz.pdf}
%\includegraphics[width=1.1\columnwidth]{polariz_numbercounts_95_wtucci_newformat.pdf}
\includegraphics[width=1.0\columnwidth]{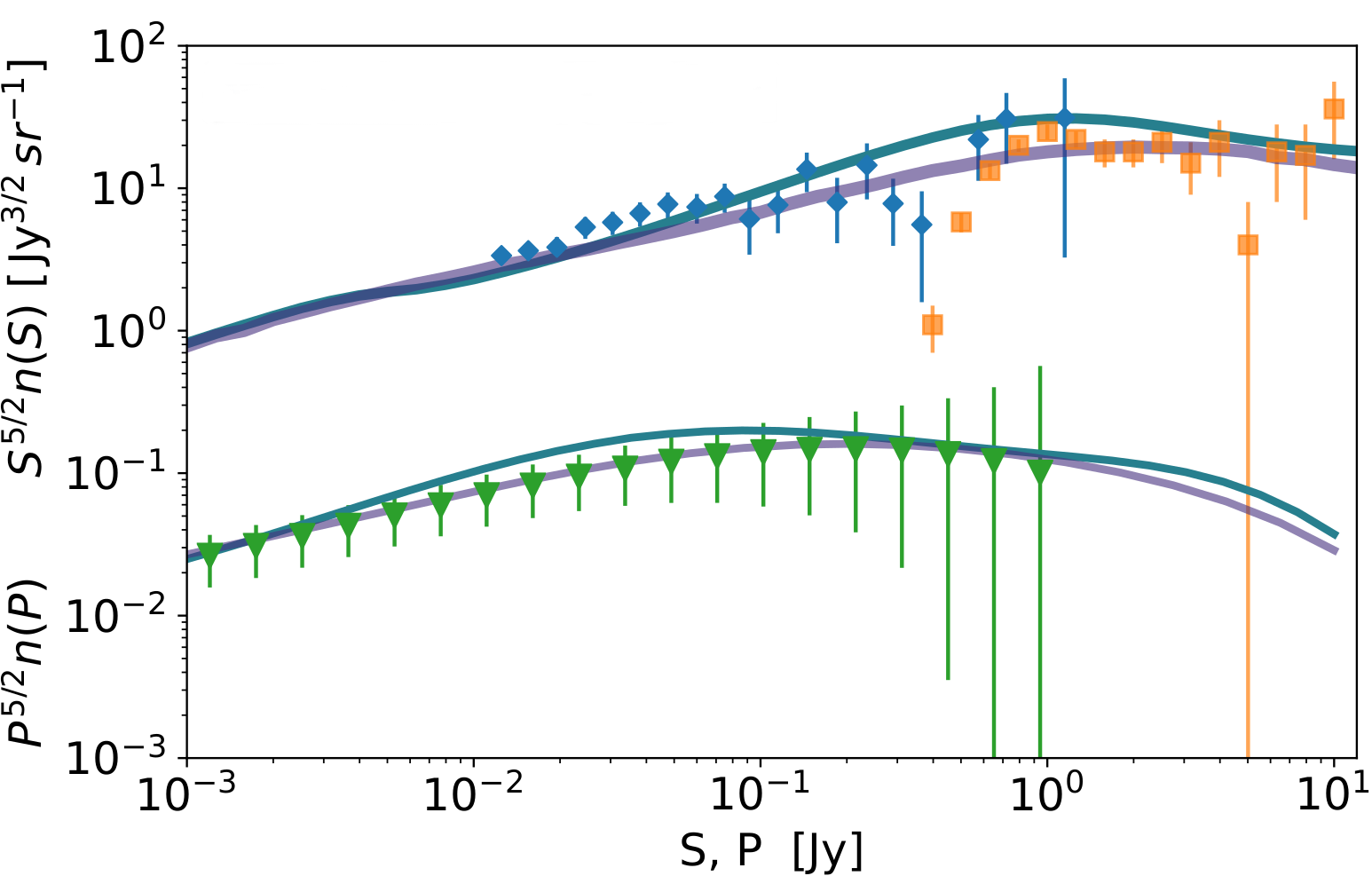}
\caption{Euclidean normalized differential number counts at $95\,$GHz. The blue lines represent the number counts
as predicted by \citet[][D05]{DeZotti2005} model in total intensity (thicker upper line) and in polarization (once convolved with our histogram of the polarization fraction, see the thinner lower line). The C2Ex model in total intensity and relative ``optimistic'' prediction in polarization presented by \citet{Tucci2012} are shown as violet lines (the thick and the thin ones, respectively). The observed total intensity source counts from SPT \citep[blue diamonds;][]{Mocanu2013} and from \textit{Planck} \citep[orange squares;][]{PlanckEarlyXIII} are also plotted. Our differential number counts in the polarized flux density computed via eq.~(\ref{equ:difsoucouP}) are shown by green triangles.}
\label{fig:SouCouP100GHz}
\end{figure}

The Euclidean-normalized differential source counts in polarized flux density
derived from eq.~(\ref{equ:difsoucouP}) down to $\simeq 1\,$mJy (approximately
the $3\,\sigma$ detection limit of our ALMA observations) are shown in Fig.~\ref{fig:SouCouP100GHz} (triangles) and listed in
Table~\ref{tab:SouCouP100GHz}. Given the relative smallness of the sample we have not distinguished among the
sub-populations considered by the \citet{Tucci2012} model (FSRQs, BL Lacs and steep-spetrum radio sources, i.e. SSRSs): the distribution of
eq.~(\ref{eq:lognorm}) was applied to all sub-populations. The error bar estimation of each data point takes into account the Poissonian contribution \citep[\textit{cf.}][]{Gehrels1986} and the uncertainties on the parameters of the lognormal distribution. To evaluate this contribution we use the semidispersion in the polarization number counts resulting from the convolution with the maximum and minimum lognormal fitting curves, respectively.
%A CHE MODELLO IN POLARIZZAZIONE SI RIFERISCONO LE LINEE?

In Fig.~\ref{fig:SouCouP100GHz} we also show, for comparison, the counts in total flux density at $100\,$GHz given by the \citet{DeZotti2005} model (``D05'', indicated by the thick blue line). The C2Ex model is displayed as a thick violet line. The observed counts are from the South Pole Telescope \citep[SPT;][]{Mocanu2013} and from Planck \citep{PlanckEarlyXIII}. In polarization we also plot the ``optimistic'' prediction for polarized source counts by \citet{Tucci2012} as a thin violet line and the convolution of the D05 model with our distribution for the polarization fraction (at $97.5\,$GHz) as thin blue line. Since the latter model tends to overestimate the source counts at such high frequency, we can assume the associated line as a ``pessimistic'' prediction. On the contrary, as displayed in Fig.~\ref{fig:SouCouP100GHz}, there is a very remarkable agreement between our current data and the model predictions by \citet{Tucci2012}.  

% %-----------------------------Figure Start---------------------------
% %\begin{landscape}
% \begin{figure*}
% \vspace{-2.5cm}
% \centering
% \includegraphics[scale=0.90]{SouCouP_100GHz.png}
% \vspace{-0.85cm}
% \caption{Normalized differential source counts at $100\,$GHz obtained by the convolution of the \cite{Tucci2011} model with the distribution for the polarization fraction derived from ALMA data.}
% \label{fig:SouCouP100GHz}
% \end{figure*}
% %\end{landscape}
% %-----------------------------Figure End------------------------------

%-----------------------------Figure Start---------------------------
%\begin{landscape}
%\end{landscape}
%-----------------------------Figure End------------------------------

\begin{table}
\caption{Euclidean normalized differential source counts at $95\,$GHz in polarized flux density
given by eq.~(\ref{equ:difsoucouP}).}
\label{tab:SouCouP100GHz}
\centering
\begin{tabular}{cccc}
\hline
$\log{[P({Jy})]}$ & $P^{5/2}n(P)$  (Jy$^{3/2}$sr$^{-1}$)& lower & upper \\
& & error & error\\
\hline
-2.920 & 0.0263 & 0.0106 & 0.0106  \\
-2.759 & 0.0308 & 0.0124 & 0.0124  \\
-2.598 & 0.0362 & 0.0145 & 0.0145  \\
-2.437 & 0.0427 & 0.0169 & 0.0169  \\
-2.276 & 0.0504 & 0.0198 & 0.0198  \\
-2.115 & 0.0594 & 0.0234 & 0.0234  \\
-1.955 & 0.0699 & 0.0278 & 0.0278  \\
-1.794 & 0.0817 & 0.0334 & 0.0334  \\
-1.633 & 0.0944 & 0.0402 & 0.0402  \\
-1.472 & 0.1075 & 0.0485 & 0.0485  \\
-1.311 & 0.1201 & 0.0583 & 0.0583  \\
-1.151 & 0.1313 & 0.0694 & 0.0694  \\
-0.990 & 0.1401 & 0.0818 & 0.0851  \\
-0.829 & 0.1456 & 0.0951 & 0.1021  \\
-0.668 & 0.1472 & 0.1088 & 0.1236  \\
-0.507 & 0.1442 & 0.1226 & 0.1544  \\
-0.346 & 0.1359 & 0.1324 & 0.1997  \\
-0.186 & 0.1214 & 0.1386 & 0.2798  \\
-0.025 & 0.1007 & 0.1911 & 0.4645  \\
%0.136 & $<$0.34383 & & \\
%0.297 & $<$0.59922 & & \\
%0.458 & $<$1.04429 & & \\
%0.619 & $<$1.81994 & & \\
\hline
\end{tabular}
\end{table}

\section{Peculiar objects}\label{sec:pecobj}

%%%%%%%%%%%%%%%%%%

Figure~9 shows, from left to right, the Stokes $I$, $Q$ and
$U$ images for the first $4$ sources, ordered in RA. Whenever the source was detected both in $Q$ and in $U$, we have superimposed to the $I$ image a vector showing the direction of linear polarization.

%%%%%%%%%%%%%%%

\setcounter{figure}{8}
\begin{figure*}
\begin{tabular}{@{\hspace{-0.75cm}}c@{\hspace{-0.4cm}}c@{\hspace{-0.4cm}}c@{\hspace{-0.0cm}}c}
\includegraphics[width=0.40\textwidth, trim={0.5cm 0.55cm 0.5cm 0.75cm},clip]{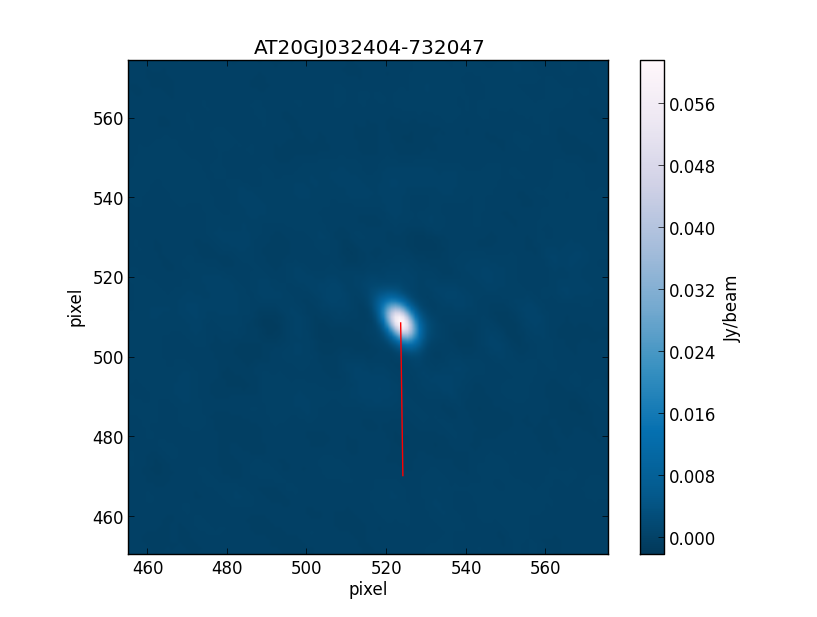}
&
\includegraphics[width=0.32\textwidth, trim={3.75cm 2.5cm 0 0},clip]{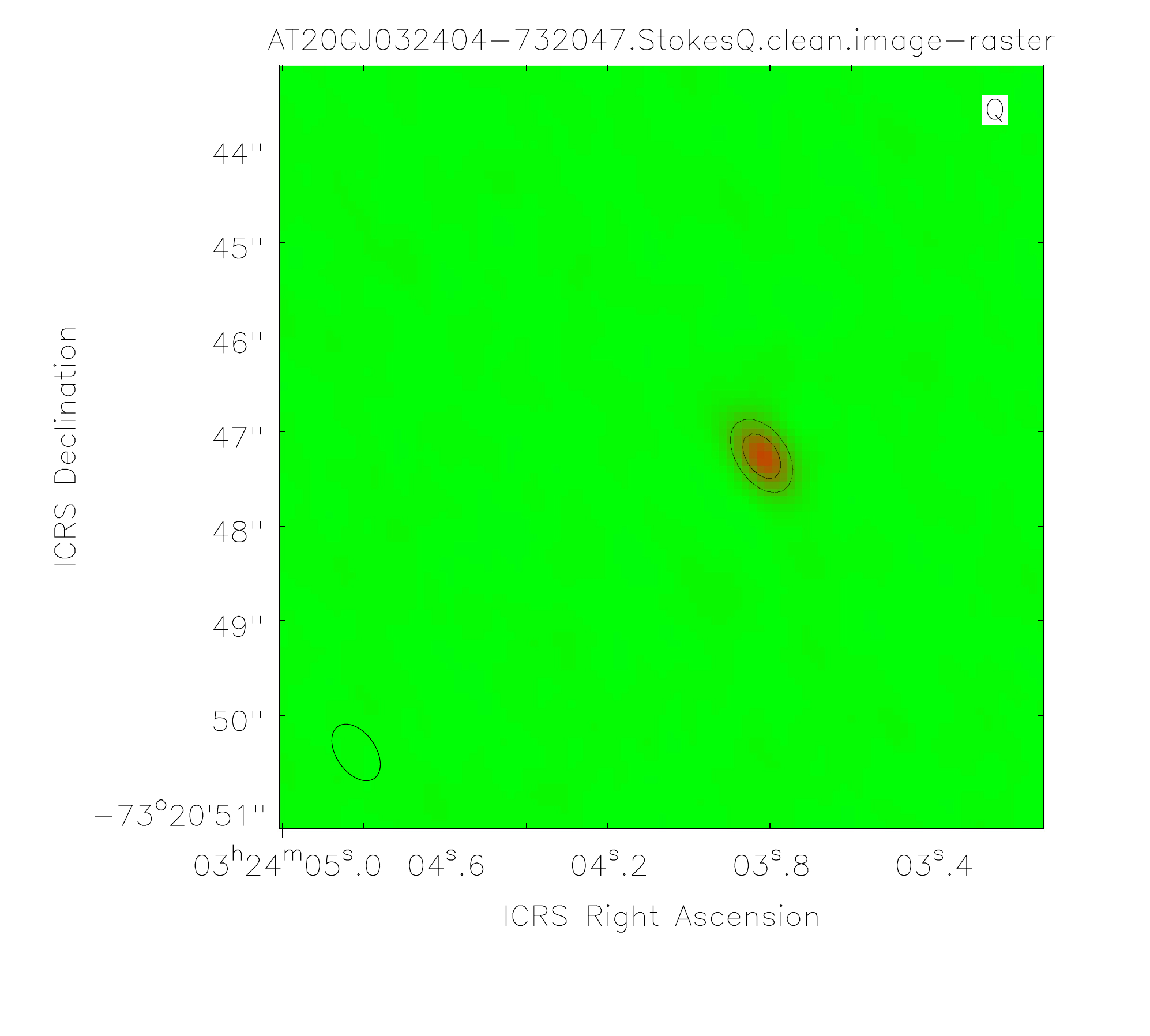}
&
\includegraphics[width=0.32\textwidth, trim={3.75cm 2.5cm 0 0},clip]{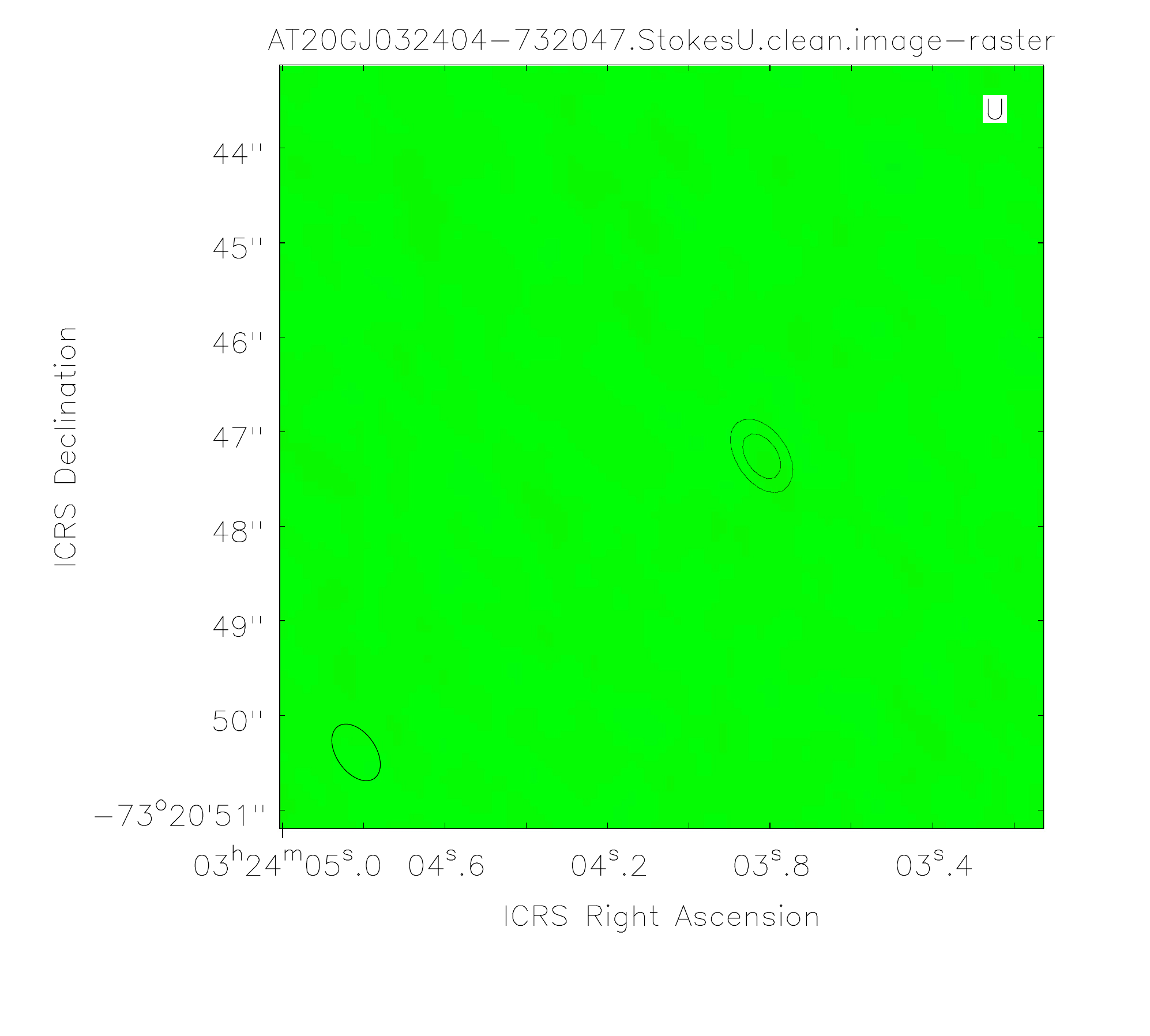}
&
\includegraphics[scale=0.22, trim={0 -0.3 0.25cm 0},clip]{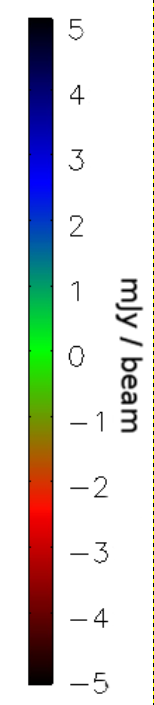}\\
\end{tabular}
\begin{tabular}{@{\hspace{-0.75cm}}c@{\hspace{-0.4cm}}c@{\hspace{-0.0cm}}c}
\includegraphics[width=0.40\textwidth, trim={0.5cm 0.55cm 0.5cm 0},clip]{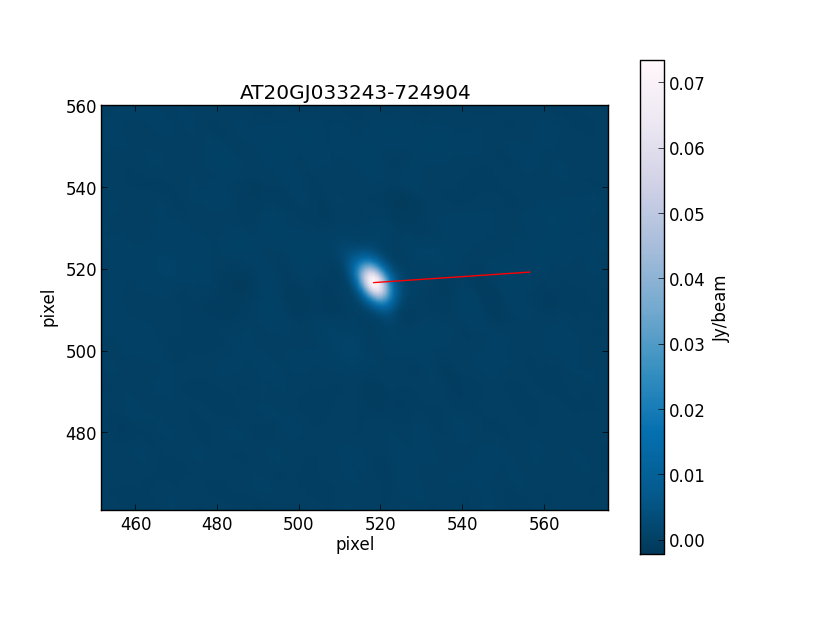}
&
\includegraphics[width=0.32\textwidth, trim={3.75cm 2.5cm 0 0},clip]{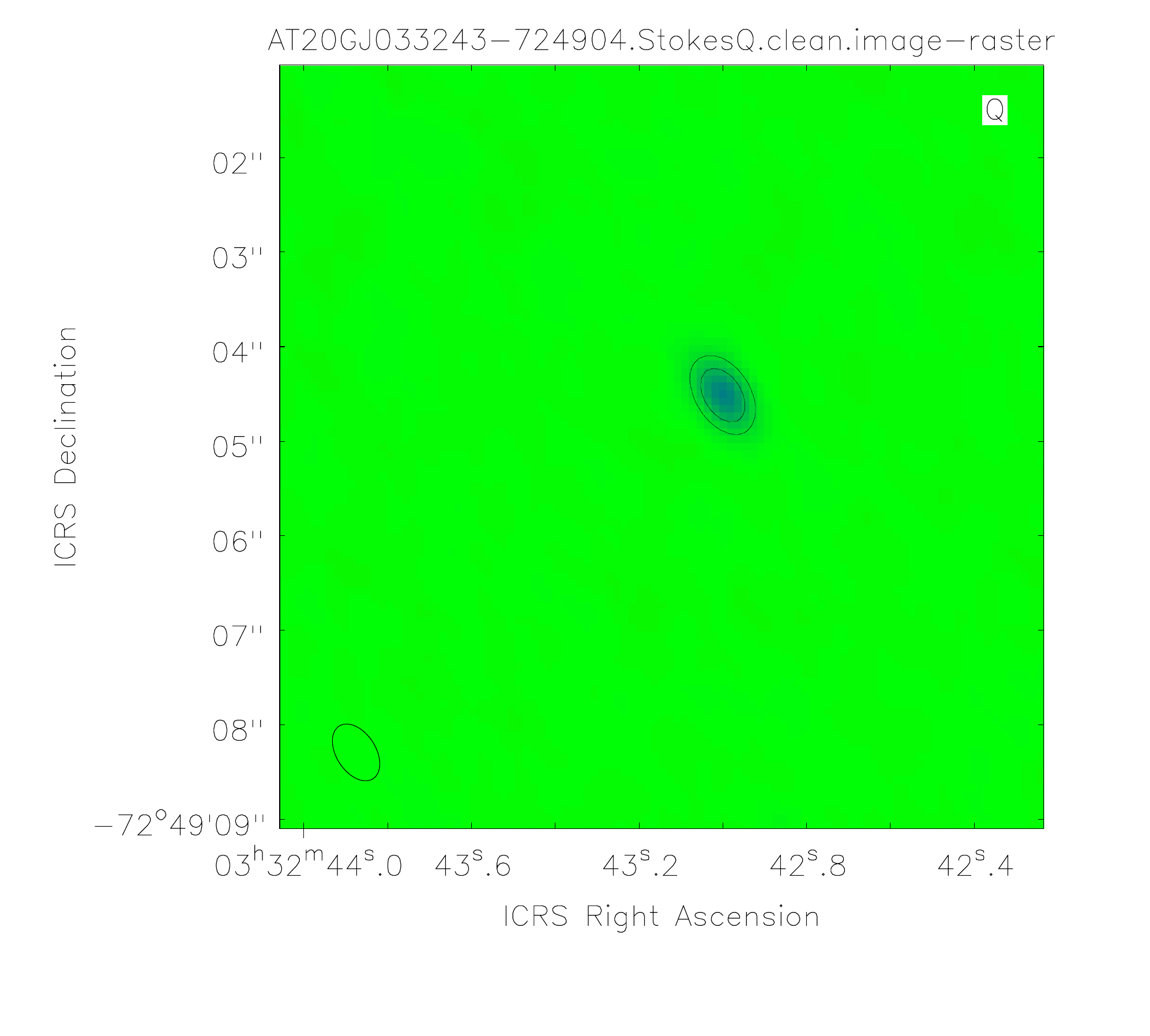}
&
\includegraphics[width=0.32\textwidth, trim={3.75cm 2.5cm 0 0},clip]{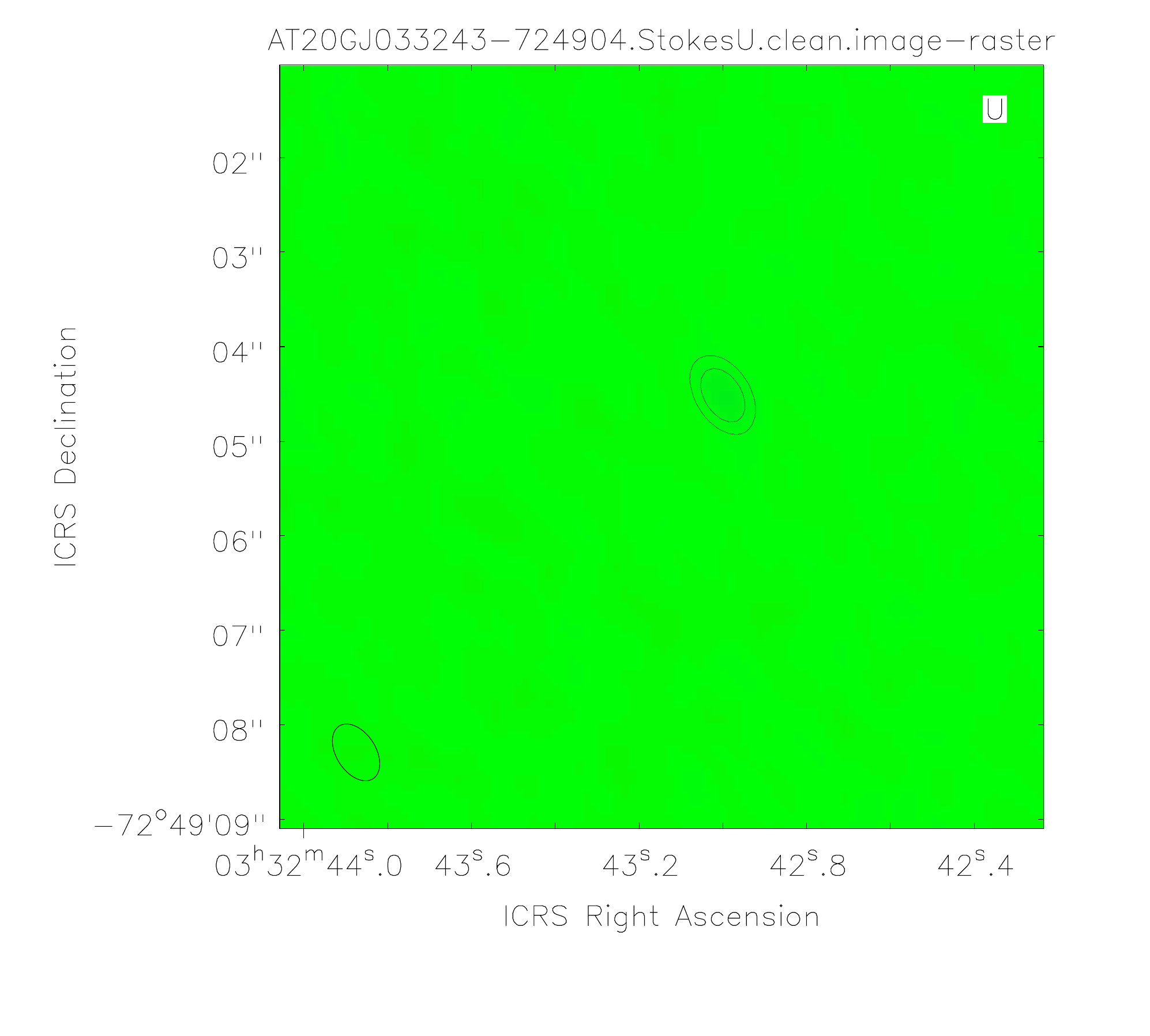}\\
\includegraphics[width=0.40\textwidth, trim={0.5cm 0.55cm 0.5cm 0},clip]{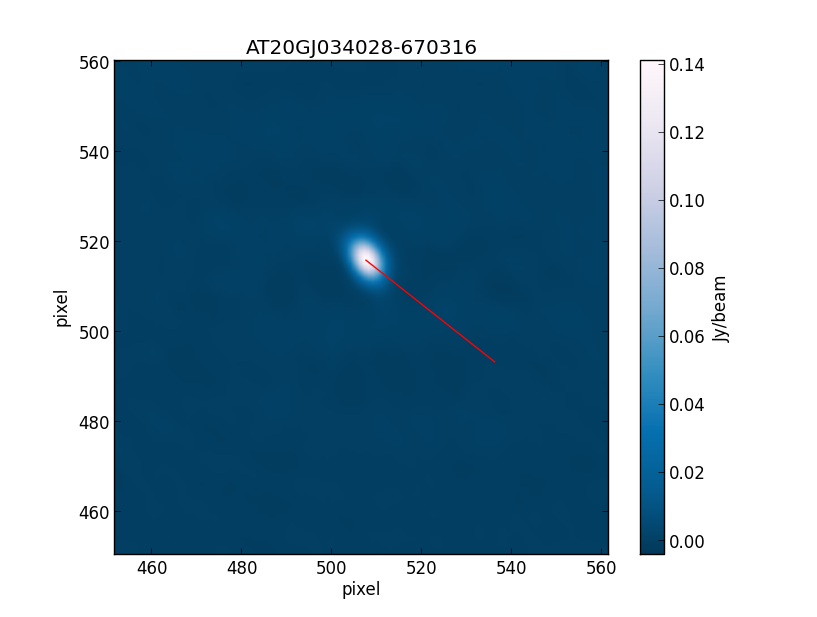}
&
\includegraphics[width=0.32\textwidth, trim={3.75cm 2.5cm 0 0},clip]{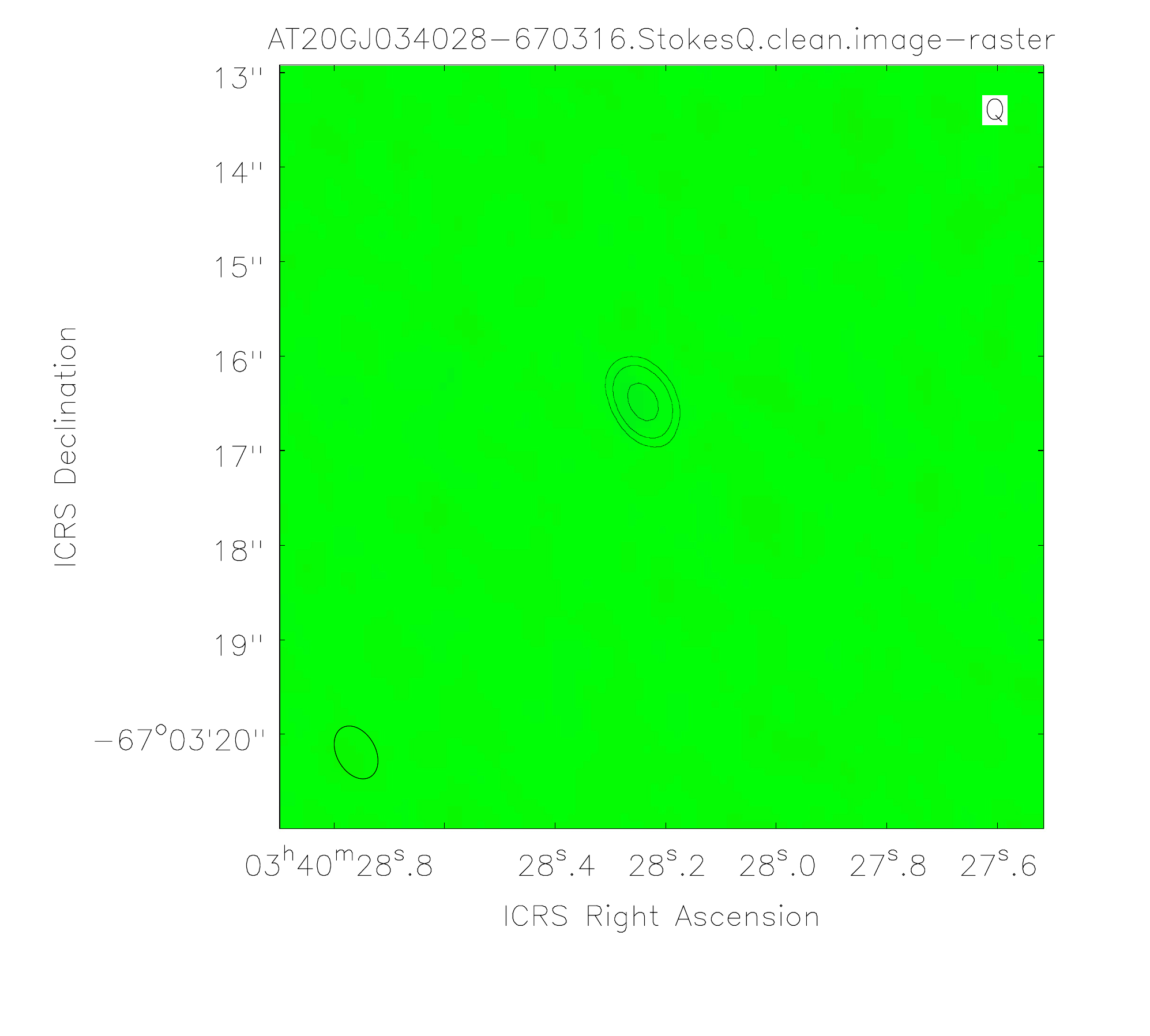}
&
\includegraphics[width=0.32\textwidth, trim={3.75cm 2.5cm 0 0},clip]{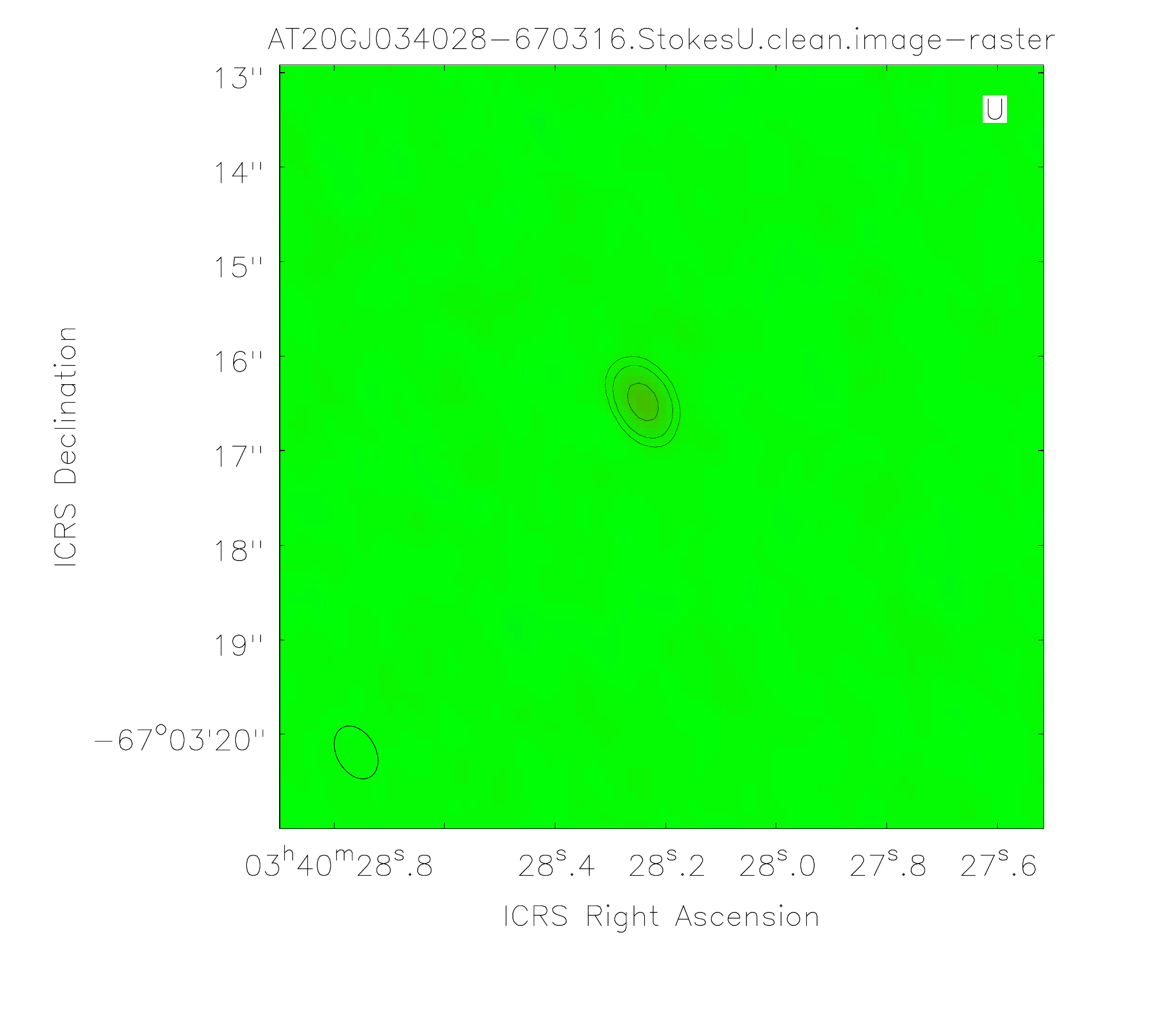}\\
\includegraphics[width=0.40\textwidth, trim={0.5cm 0.55cm 0.5cm 0},clip]{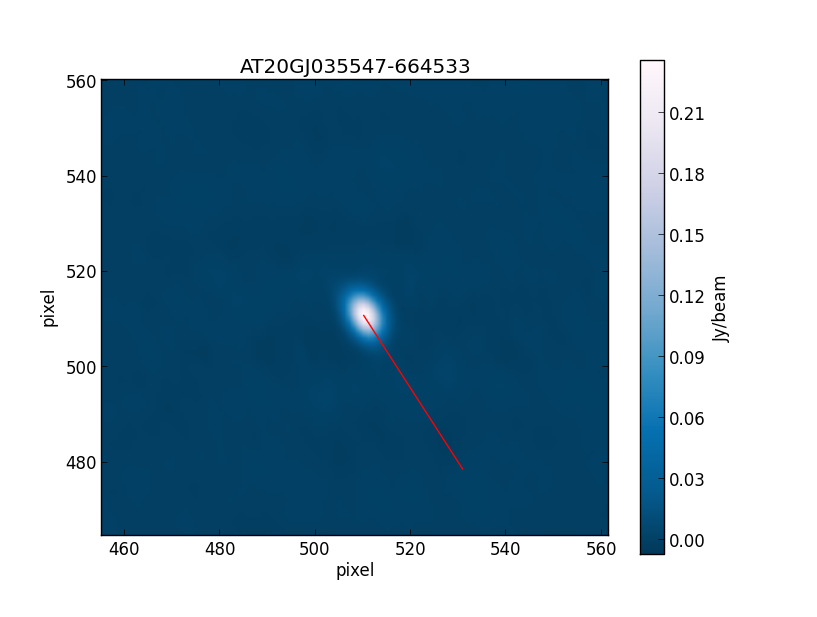}
&
\includegraphics[width=0.32\textwidth, trim={3.75cm 2.5cm 0 0},clip]{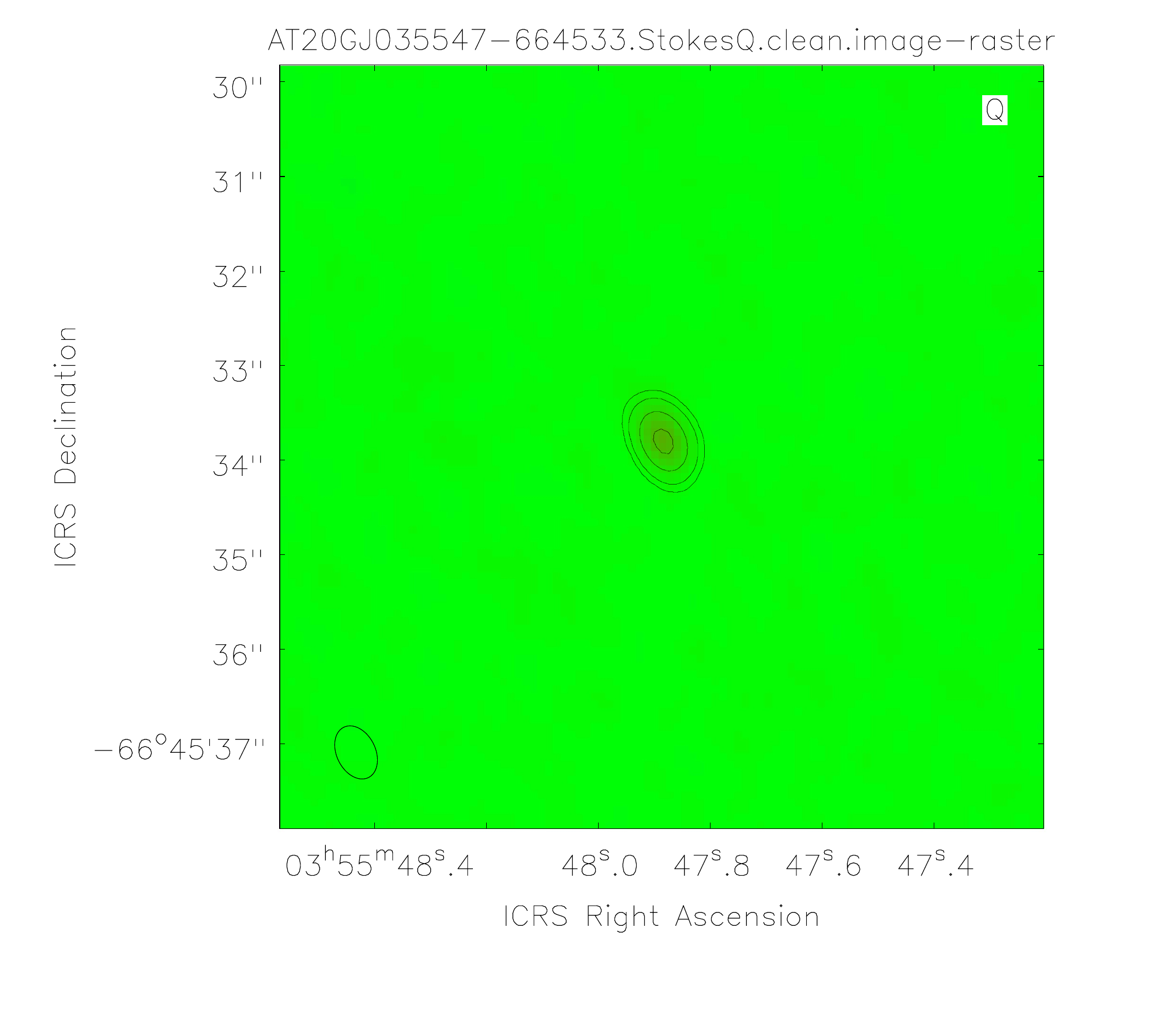}
&
\includegraphics[width=0.32\textwidth, trim={3.75cm 2.5cm 0 0},clip]{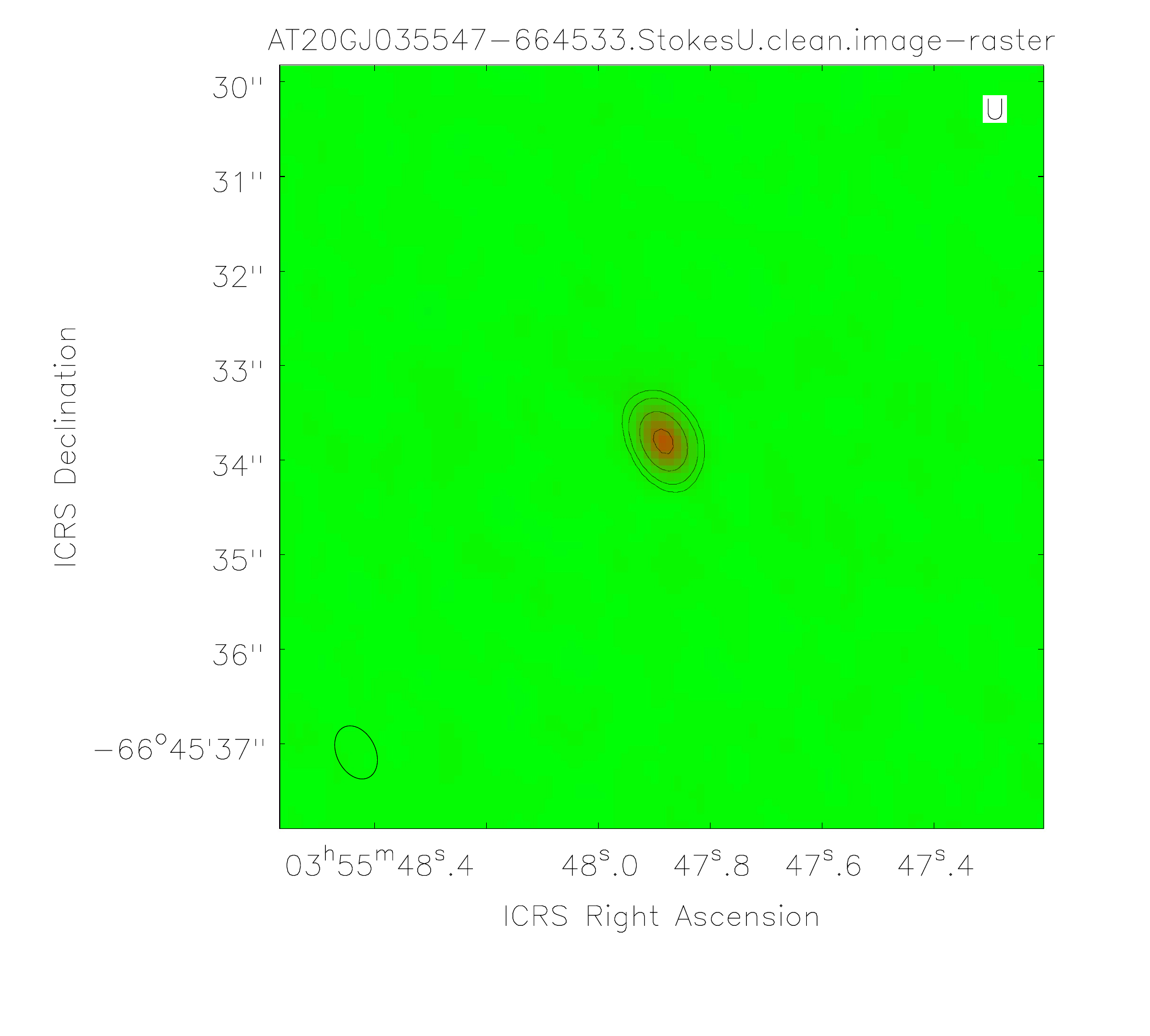}\\
\end{tabular}
\label{fig:ALMAIQU_Images}
\vspace{-0.25cm}
\caption{Stokes $I$, $Q$ and $U$ images of the first $4$ objects (ordered by RA) observed
with ALMA. We have superimposed to each $I$ map polarization vectors indicating the
polarization angle, by using the Key analysis Automated FITS-Image Explorer \citep[KAFE;][]{Burkutean2018} package. For $Q$ and $U$ maps we use the same data range (displayed by the colorwedge attached to the first U map). We also overplot total intensity maps by contours with levels at $20$, $50$, $100$, $200$, $500$ and $1000\,$mJy. The images for the other objects are provided online as supplementary material.}
\end{figure*}

%-----------------------------Figure Start---------------------------
%\begin{landscape}
\begin{figure*}
%\vspace{-2.5cm}
\centering
\includegraphics[width=1.0\textwidth]{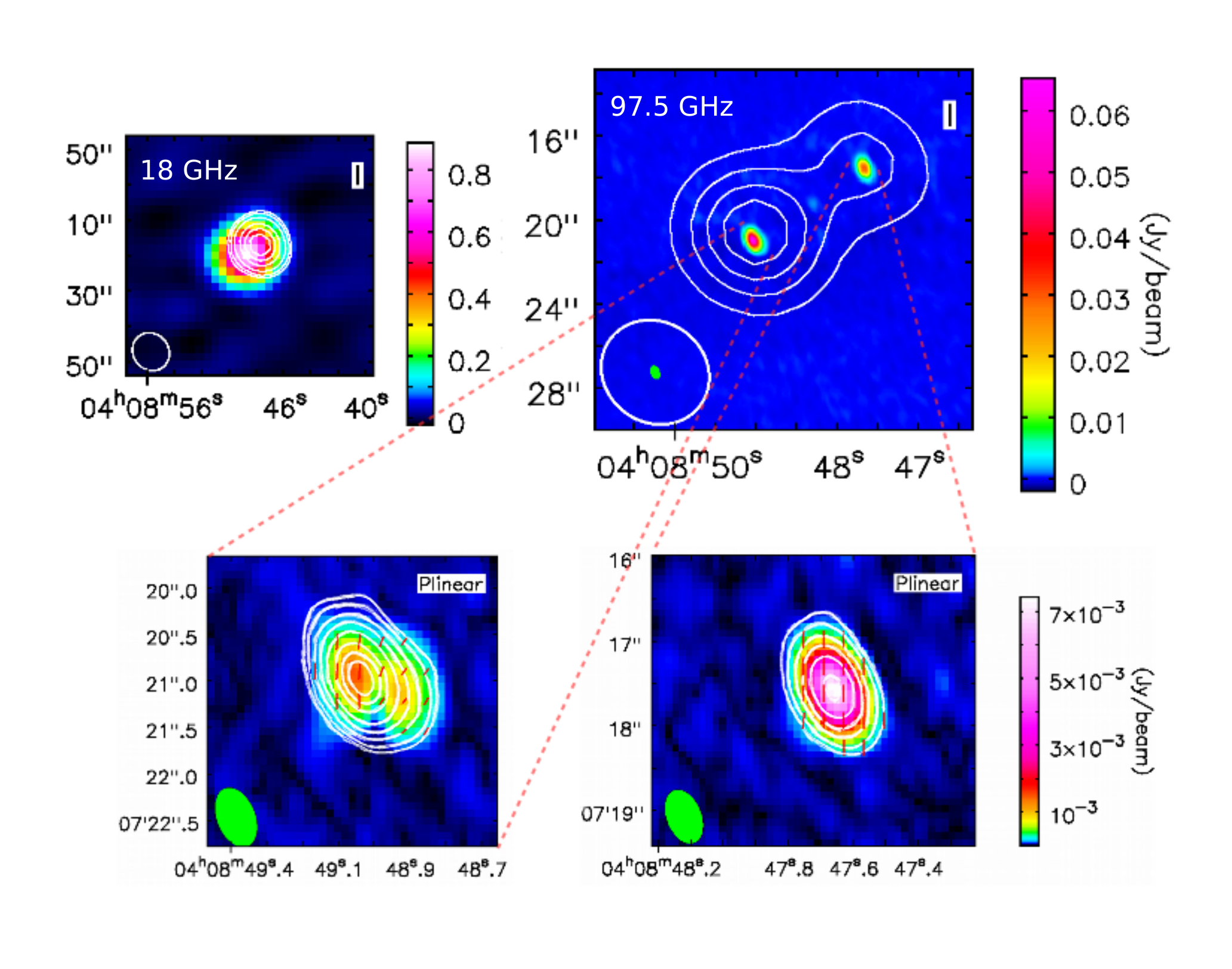}
%\vspace{-0.85cm}
\caption{Images of AT20GJ040848-750720. The upper-left panel shows the ATCA image at $18\,$GHz, where the white contours refer to polarization, the other multi-colour contours refer to total intensity. The upper-right panel shows the total intensity ALMA image at $97.5\,$GHz (multi-colour contours) together with total intensity contours (more extended lines) from the ATCA image at $38\,$GHz. In the bottom left corner of each panel we display the ATCA (white contour) and ALMA (green filled ellipse) synthesized beams.
The lower panels show the ALMA polarization image (multi-colour) and total intensity (white) contours with superimposed
the magnetic field directions (in red) for the eastern and the western hotspots (left and right panel, respectively).}
\label{fig:AT20GJ040848-750720_composed}
\end{figure*}
%\end{landscape}
%-----------------------------Figure End------------------------------
All the objects in the sample were selected as being point-like at $20\,$GHz
but some of them are spatially resolved by ALMA. Sources
AT20GJ040820-654508, AT20GJ050644-610941, AT20GJ063546-751616, AT20GJ074331-672625 and AT20GJ080633-711217 seem to display jet components displaced from the central core. However, such
components are at least two orders of magnitude fainter than the core (see
images similar to Fig.~9 provided as supplementary online material); hence, for the purposes of this paper, these
sources are effectively point-like.

Instead AT20GJ040848-750720 and the leakage calibrator, PKS0521-365 (which,
however, does not belong to the sample), are well resolved by ALMA and show a
peculiar structure in polarization. Therefore they deserve more discussion.

{\bf AT20GJ040848-750720}. This is an FR-II source at $z\simeq 0.69$. It was
unresolved by ATCA at $20\,$GHz, although the centroid in polarization was
slightly offset from that in total intensity. The ALMA image (with
$0.5\,$arcsec resolution) shows that the emission is dominated by two bright
lobes ({\it cf.} Fig.~\ref{fig:AT20GJ040848-750720_composed}). Both exhibit a high
depolarization, slightly higher in the eastern one. The latter also shows a double
structure in the polarized emission. The core sits midway of the two
lobes and is quite faint, i.e. $\simeq 1.5\,$mJy ({\it cf.} Fig.~\ref{fig:AT20GJ040848-750720_composed}).
%The ALMA image  also reveals a double substructure in the polarized intensity of the eastern lobe.

\begin{figure}
%\vspace{-2.5cm}
\centering
\includegraphics[width=1.05\columnwidth,trim={5.15cm 2cm 0 0},clip]{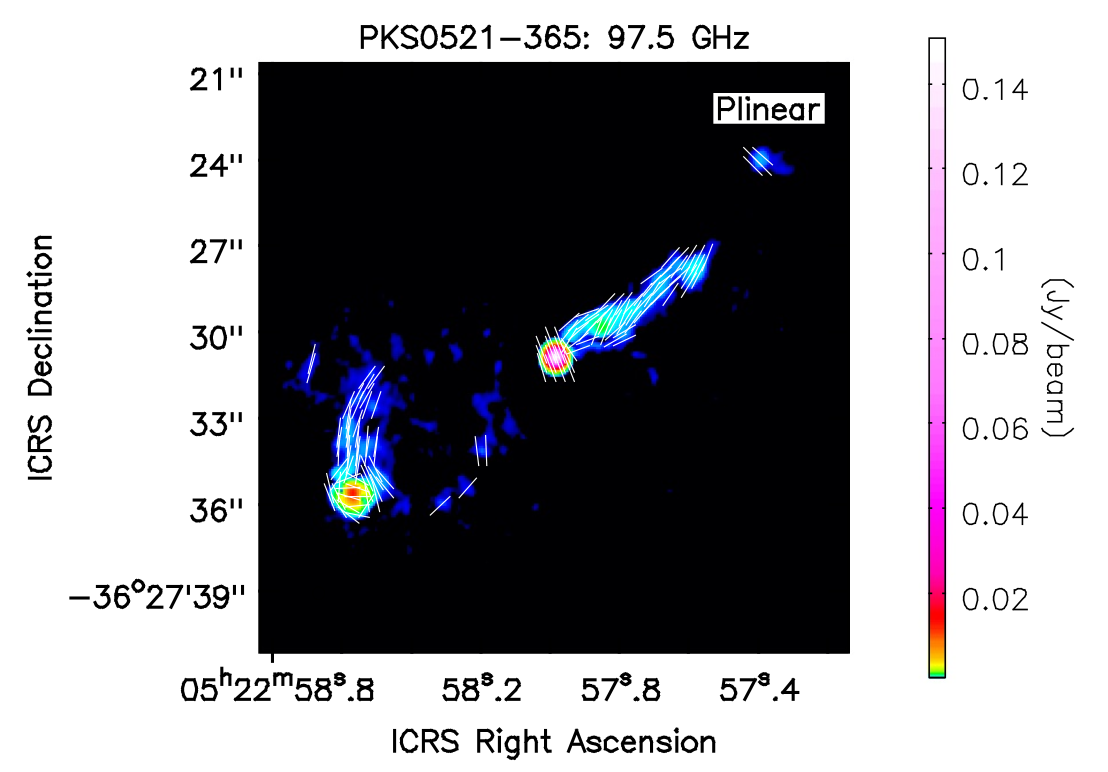}
%\vspace{-0.85cm}
\caption{Linearly	polarized emission (multi-colour contours) and	position angles (vectors) in ALMA Band 3 of PKS0521-365.}
\label{fig:PKS0521-365}
\end{figure}

{\bf PKS0521-365}. This nearby ($z=0.0554$)	radio-loud	object is	a bright
$\gamma$-ray source and exhibits a	variety	 of	nuclear	and	extranuclear
phenomena \citep{Falomo2009}. It is one of	the	most remarkable	objects in	
the	southern sky: it	is one of	the	three	known	BL Lac objects	showing	
a	kiloparsec-scale	jet	well	resolved	at	all	bands	
\citep{Liuzzo2011}.	The ALMA image (Fig.~\ref{fig:PKS0521-365}) shows a
one-sided radio jet extending in the N-W direction up	to	$7\,$arcsec from
the nucleus. The jet exhibits many knots,	also	detected	from	the
optical	to	X-rays 	\citep{Falomo2009}.	A hotspot located at	$8\,$arcsec from	
the	nucleus	in	the	south-east direction	is	also	detected in	all	bands.	
At	low	frequency,	the	arcsecond-scale	radio	structure	 is	dominated	by	
an	extended	lobe.	The	overall	energy	distribution	of	PKS	0521-365	
is	consistent	with	a	jet	oriented	at	about	$30^\circ$ with	respect	
to	the	line	of	sight.	This	is	also	in	agreement	with	the	
absence	of	superluminal	motion	in	the	parsec-scale	jet
\citep{Falomo2009}. In the millimeter	bands,	 extended	structures	
(hotspot	 and	 jet)	 of this	 object	are detected up	to $320\,$GHz;
their morphology is similar to that observed from the	optical	to X-rays	
\citep{Liuzzo2015, Leon2016}.
% \citet{Liuzzo2015} presented an	estimate of	molecular	gas	content	together with	an analysis	of the	SED	of each	source component.		

Polarimetric data for such resolved objects is very helpful to perform studies
aimed at addressing fundamental questions about the AGN physics, such as the
role of the magnetic field in jetted/radio loud AGNs, the plasma properties and
particle acceleration mechanisms. By using more advanced techniques, such as
the Faraday Rotation (FR) Synthesis \citep{Brentjens2005} or procedures similar
to those adopted by \citet{O'Sullivan2012}, it is possible to obtain a 3D
representation of the magnetic field. A paper from our collaboration (Liuzzo et
al., in preparation) will exploit such techniques on PKS0521-365 maps, trying,
among other things, to address the physical processes operating in the hotspots
\citep[e.g., Fermi-II acceleration or multiple shocks, {\it cf.}][]{Prieto2002}.

\subsection{Calibrator candidates for CPR studies}
As briefly discussed in the Introduction, the Cosmic Polarization Rotation (CPR) studies, which rely on the statistical analysis of a collection of objects, typically suffer from the lack of zero point calibration of the polarization angle. Thus, having reference objects whose polarization angle is known on a sub-degree scale is particularly helpful for those studies: good candidates might be compact radio sources which show bright total flux densities (at least of few hundreds of mJy), with a polarization fraction at least of a few $\%$ and a reasonably stable behaviour in the polarization angle. Until now, the essential lack of polarimetric observations at high frequencies and consequent monitoring on large samples of radio sources make these calibrators very rare, especially in the southern sky.

Both our ALMA and ATCA polarization angle measurements are free from the zero point systematic error arising from the phase difference in the cross-correlation products of the reference antenna. In the case of ATCA each antenna receiver at the frequencies we observed is equipped with a noise diode mounted in one of the linear feeds. The signal injected by each diode is received by the other feed and phase differences are characterized for all the antennas and stored in visibility files. During the data reduction with MIRIAD (the standard radio interferometry package for ATCA) a reference antenna is set and the relative phase difference correction is applied to the data. In the case of ALMA, the determination of this systematic term can be achieved by observing a polarized object at different parallactic angles for a typical angular coverage of at least $3\,$hr, in order to break degeneracies associated to the unknown polarization signal of the calibrator itself and  leakage terms. Thus, we searched in our sample (up to $105.5\,$GHz) sources suitable as CPR calibrators. 
%\FloatBarrier
%\vspace{-10cm}
%-----------------------------Figure Start---------------------------
%\begin{landscape}
%\end{landscape}

%-----------------------------Figure End------------------------------
%\FloatBarrier
We firstly restricted ourselves to those found to be the less variable ones in both total intensity and polarization (typically less than $10\%$) and which are also stable in the polarization angle at the different epochs we observed. Then, we selected those with relatively high flux densities (at least $100\,$mJy at $97.5\,$GHz) and polarization fractions (especially at frequencies higher than $20\,$GHz) at least at a few $\%$ level. The first object we selected is PKS0637-752 (also known as AT20GJ063546-751616), already suggested by \citet{Massardi2013} as a potential leakage calibrator, being at $\sim 1\,$Jy and $\sim 1.6\%$ polarized at $97.5\,$GHz. Our ATCA observations show that the polarization angle is quite constant across the $5.5-38\,$GHz frequency range and stable within $8^\circ$ at $38\,$GHz between 2014 September and 2016 July (see Fig.~\ref{fig:Spettri1}). However, this object displays Faraday rotation in different frequency regimes (see Table~\ref{tab:ALMARMdetlist}): at ALMA frequencies, between $90$ and $105\,$GHz the polarization angle absolute variation is $\simeq 14^\circ$. Other somewhat fainter but more polarized objects we found in our sample are AT20GJ062005-610732 ($120\,$mJy, $10.5\%$ polarized) and AT20GJ074331-672625 ($190\,$mJy, $5.2\%$ polarized): the first one is constant within $8^\circ$ both across the $18-38\,$GHz frequency range and between the two epochs; the second is less constant between the different frequencies but stable within $\sim 3^\circ$ both at $33$ and $38\,$GHz.

The objects we have identified are potential calibrators for CPR studies: the fact their variability in the polarization angle is no more than $8^\circ$, over a period of almost $2$ years, might indicate a stability at sub-degree level over a period of (at least) few days. The latter is the main requirement to reduce systematics in CPR experiments as well as for CMB studies. We are going to monitor on a more regular basis these objects both with ATCA and with ALMA at higher frequencies (Band 3 and 6, i.e. $97.5$ and $243\,$GHz, respectively).

\section{Conclusions}
\label{sec:discusseconcl}

We have presented and discussed high sensitivity ALMA polarimetric observations
in Band $3$ ($97.5\,$GHz) of a complete sample of $32$ extragalactic radio
sources (in the region with $b< -75^\circ$) drawn from the faint PACO sample, i.e. compact AT20G sources with $S_{20\rm GHz}\ge 200\,$mJy. The
rms in polarized flux density was $0.4\,$mJy, which allowed a
detection rate of $94\%$ at $5\,\sigma$.
%We have presented and discussed high sensitivity polarimetric observations in $7$ bands, centered at $2.1$, $5.5$, $9$, $18$, $24$, $33$ and $38\,$GHz, of a complete sample of 104 extragalactic sources with $S_{20\rm GHz}\ge 200\,$mJy in the AT20G catalogue. The r.m.s error in the polarized flux density is $0.6\,$mJy at $\nu \ge 5.5\,$GHz and slightly higher ($1\,$mJy) at $2.1\,$GHz due to the heavy rfis contamination.

ALMA observations (together with ATCA and GLEAM data) allowed us to reach more
than $3$ decades of spectral coverage in total intensity and $\sim 1.7$ decades
in polarization. Most of the sources (26 out of 32) revealed a flux density excess in total intensity with
respect to spectra extrapolated from ATCA data at lower frequencies (collected
between $2$ and $6$ months before ALMA measurements), suggesting the emergence
of another emission component. The high frequency emissions are polarized at
a few percent level. None of the observed spectra showed signs of any synchrotron break,
and the spectral indices in total intensity between $36.5$ and $97.5\,$GHz are typically flat,
i.e. $\simeq -0.19$. %{\bf Very remarkably, this result on the median spectral indices is in perfect agreement with the predictions of \citet{Tucci2011} (C2Co model, see their Table~6), albeit the frequency interval is slightly different (i.e. $30-100\,$GHz).}

The distribution of polarization fractions observed with ALMA allowed us to extend the analysis of \citet{Galluzzi2018} up to
$97.5\,$GHz, confirming the absence of any statistically significant trend with
the frequency (or the flux density). This data set has been included in the analysis described in \citet{Puglisi2018}, which presents the state-of-the-art about polarimetry of extragalactic radio sources and provides forecasts for their contamination of the B-mode angular power spectrum, useful for current and forthcoming CMB experiments. Besides, our observed polarization fractions further confirm the results obtained from Planck maps by \citet[][adopting a stacking technique]{Bonavera2017} and by \citet[][exploiting the intensity distribution analysis,
IDA, method]{Trombetti2018}.

We also looked for differences in the high frequency polarization properties of
different sub-classes of sources, using classifications based on spectral
indices or on the number of components detected in source spectra, but the
smallness of the sample prevented any firm conclusion.
%Polarization measurements in the range $5.5-38\,$ GHz for $53$ objects of the sample were reported by \citet{Galluzzi2017}. The measurements for the other $51$ sources are new, as are the $2.1\,$GHz measurements for the full sample of $104$ sources. The $53$ sources were re-observed at $5.5$ and $9\,$GHz, while we managed to repeat the observations at $18$, $24$, $33$ and $38\,$GHz only for $20\%$ of them. The previous measurements at $33$ and $38\,$GHz were re-calibrated using the updated model for the flux density absolute calibrator, PKS1934-638, that was not available for the earlier analysis but is now loaded into the MIRIAD package.

By exploiting the $8\,$GHz ALMA bandwidth, we investigated the RMs at $\sim
100\,$GHz. We found intrinsic values $\simeq
6.4\times 10^4\,\hbox{rad}\,\hbox{m}^{-2}$, at least one order of
magnitude higher than those obtained for the $18-38\,$GHz frequency range and
two orders of magnitude higher than in the  $2-9\,$GHz range. Although with
large uncertainties, these results suggest the presence of dense screens of
magnetized plasma that can strongly depolarize the mm-wave emission,
suppressing the increase in the polarization fraction due to more ordered
magnetic fields, typically expected in the regions of the jet closer to the nucleus.

We have also presented estimates of source counts in linearly polarized flux
density at $95\,$GHz, derived from the convolution of the model C2Ex by
\citet{Tucci2011} for total intensity source counts with the distribution of
polarization fractions for our sample.
% This result has been exploited by \citet{Puglisi2018} to estimate the contamination of the Cosmic Microwave Background angular power spectrum by unresolved radio sources.

Two objects in our dataset, namely the target AT20GJ040848-750720 and the calibrator PKS0521-365, show well-resolved structures, which constitute interesting case studies to constraint magnetic fields and particle acceleration mechanisms along AGN jets and in hotspots: a preliminary description of these sources has been presented here. However, a more exhaustive investigation will be addressed by future publications (e.g. Liuzzo et al., in preparation) as well as by further observations. In fact, maps	obtained for PKS0521-365 show that ALMA, with an angular resolution $\sim 0.2\,$arcsec, can reveal	polarized	emission even	in the lobes, by spending only $10\,$min on source. This demonstrates 	
the power of ALMA in detecting also	faint ($<0.1\,$mJy)	source	components for large samples of sources.

Finally, by considering the less variable but (at the same time) the brightest and the most polarized objects in our sample, we have identified three cases that display particular stability in the polarization angle, both in time and frequency (especially at higher frequencies). These (as well as similar) objects may serve as polarization angle calibrators for improving future CPR studies, by reducing the currently limiting calibration error below the degree level.

\section*{Acknowledgments}
%We thank the anonymous referee for useful comments.
This paper makes use of the following ALMA data: ADS/JAO.ALMA\#2015.1.01522.S. ALMA is a partnership of ESO (representing its member states), NSF (USA) and NINS (Japan), together with NRC (Canada) and NSC and ASIAA (Taiwan) and KASI (Republic of Korea), in cooperation with the Republic of Chile. The Joint ALMA Observatory is operated by ESO, AUI/NRAO and NAOJ.
We acknowledge financial support by the Italian {\it Ministero dell'Istruzione,
Universit\`a e Ricerca}  through the grant {\it Progetti Premiali 2012-iALMA}
(CUP C52I13000140001).
%Acknowledgements from other authors
We gratefully acknowledge financial support from the INAF PRIN SKA/CTA 
project FORmation and Evolution of Cosmic STructures (FORECaST) with Future Radio Surveys.
Partial support from ASI/INAF Agreement 2014-024-R.1 for the {\it Planck} LFI
Activity of Phase E2, from the ASI/Physics Department of the university of
Roma--Tor Vergata agreement n. 2016-24-H.0  and from ASI through the contract
I-022-11-0 LSPE is acknowledged.
%Acknowledgements from other authors
We thank the staff at the Australia Telescope Compact Array site, Narrabri
(NSW), for  the valuable support they provide in running the telescope and in
data reduction. The Australia Telescope Compact Array is part of the Australia
Telescope which is funded by the Commonwealth of Australia for operation as a
National Facility managed by CSIRO. AB acknowledges support from the European
Research Council under the EC FP7 grant number 280127.
%MM and VG thank Joe Callingham for the useful discussions.
%VG also thanks Rocco Lico for the useful discussions.
VC acknowledges DustPedia, a collaborative focused research project supported
by the  European Union under the Seventh Framework Programme (2007-2013) call
(proposal no. 606824). The participating institutions are: Cardiff University,
UK; National Observatory of Athens, Greece; Ghent University, Belgium;
Université Paris Sud, France; National Institute for Astrophysics, Italy and
CEA (Paris), France. %LT and LB acknowledges financial support from the I+D 2015 project AYA2015-65887 (MINECO/FEDER).
LB and LT acknowledge the PGC 2018 project PGC2018-101948-B-I00 (MINECO/ FEDER)

\bsp

%\label{lastpage}

\end{document}